\documentclass[journal]{IEEEtran}

\usepackage{graphicx,xcolor}
\graphicspath{{../pdf/}{../jpeg/}}
\DeclareGraphicsExtensions{.pdf,.jpeg,.png}
\usepackage{amsmath}
\usepackage{amsfonts}
\usepackage{array}
\usepackage{amssymb}
\graphicspath{{results/}}
\usepackage{empheq}
\usepackage{epstopdf}
\usepackage[linesnumbered,ruled]{algorithm2e}
\newcounter{MYtempeqncnt}
\usepackage{array}
\usepackage{caption}
\usepackage{subcaption}

\begin{document}

\title{Reinforcement Learning for Adaptive Caching\\ with Dynamic Storage Pricing}
\author{Alireza Sadeghi, Fatemeh Sheikholeslami, Antonio G. Marques, and Georgios B. Giannakis
	\thanks{{Alireza Sadeghi, Fatemeh Sheikholeslami, and Georgios B. Giannakis are with the Digital Technology Center and the Department of Electrical and Computer engineering, University of Minnesota, Minneapolis, USA. \newline Emails: \{sadeghi, sheik081, georgios\}@umn.edu}
		{\newline Antonio G. Marques is with the Department of Signal Theory and Communications, King Juan Carlos University, Madrid, Spain. \newline Email: antonio.garcia.marques@urjc.es}
		{\newline The work in this paper has been supported by USA NSF grants 1423316, 1508993, 1514056, 1711471, by the Spanish MINECO grant OMICROM (TEC2013-41604-R) and by the URJC Mobility Program. Part of this work has been presented in ICASSP 2018, Calgary, Canada~\cite{ICASSPversion}.}  }}
\maketitle

\begin{abstract}
Small base stations (SBs) of fifth-generation (5G) cellular networks are envisioned to have  storage devices to locally serve requests for reusable and popular contents by \emph{caching} them at the edge of the network, close to the end users. The ultimate goal is to shift part of the predictable load on the back-haul links, from on-peak to off-peak periods, contributing to a better overall network performance and service experience. To enable the SBs with efficient \textit{fetch-cache} decision-making schemes operating in dynamic settings, this paper introduces simple but flexible generic time-varying fetching and caching costs, which are then used to formulate a constrained minimization of the  aggregate cost across files and time. Since caching decisions per time slot influence the content availability in future slots, the novel formulation for optimal fetch-cache decisions falls into the class of dynamic programming. Under this generic formulation, first by considering stationary distributions for the costs and file popularities, an efficient reinforcement learning-based solver known as value iteration algorithm can be used to solve the emerging optimization problem. Later, it is shown that practical limitations on cache capacity can be handled using a particular instance of the generic dynamic pricing formulation. Under this setting, to provide a light-weight online solver for the corresponding optimization, the well-known reinforcement learning algorithm, $Q$-learning, is employed to find optimal fetch-cache decisions. Numerical tests corroborating the merits of the proposed approach wrap up the paper. 
\end{abstract}
\begin{IEEEkeywords}
Dynamic Caching, Fetching, Dynamic Programming, Value iteration, $Q$-learning.
\end{IEEEkeywords}

\section{Introduction}
In the era of data deluge, storing ``popular'' contents at the edge of a content delivery network (CDN) or 5G cellular network, is a promising technique to satisfy the users' demand while alleviating the congestion on the back-haul links \cite{paschos2018, paschos16, Magzine}. To this aim, small basestations (SBs) equipped with a local cache must intelligently store reusable popular contents during off-peak periods, and utilize the stored data during on-peak hours. To endow SBs with the required learning capability, a wide range of learning and optimization approaches has been adopted (see \cite{paschos2018}, \cite{paschos16}).
	
Considering \textit{static} popularity for contents, a multi-armed bandit formulation of the problem was investigated in \cite{Gunduz}, where the caching is carried out according to demand history and under unknown popularities. Coded, convexified, and distributed extensions of this problem were later studied in \cite{Sengupta}, context and trend-aware learning approaches in \cite{Schaar_2017}, \cite{Schaar_trend}, and coordinated-distributed extensions in \cite{Coordinated}. From a learning perspective, the trade-off between the ``accuracy'' of learning a static popularity, and the corresponding learning ``speed'' is investigated in \cite{Trade_off} and \cite{Learningbound}.

In reality however, popularities exhibit fluctuations meaning they are \textit{dynamic} over a time horizon. For instance, half of the top 25 requested Wikipedia  articles change on a daily basis~\cite{paschos2018,Wikidynamic}. This motivates  recent approaches to designing caching strategies under dynamic popularity scenarios. 

To account for dynamic popularities, a Poisson shot noise model was introduced in \cite{PSN1}, followed by an age-based thresholding caching strategy in \cite{PSN2}. Furthermore, reinforcement learning-based approaches were studied in \cite{RL1,RL1conf} and \cite{RL2}. In \cite{RL1}, global and local popularities are modeled by different Markov processes, and a $Q$-learning based algorithm was proposed; while in \cite{RL2}, a policy gradient approach was followed to optimize a parametric policy. From an ``accuracy-speed'' trade-off perspective, a class of learning-based algorithms under dynamic popularities was analyzed in~\cite{GunduzAccspeed}. Modeling the evolution of the popularities as Markov processes, an online coded caching scheme was introduced in~\cite{onlineCodedCaching}, to minimize the long-term average transmitted data over the back-haul. Likewise, delivery time was minimized in \cite{ITDeniz} through a coded caching strategy. 

Targeting different objectives, optimization-based dynamic caching has been utilized in different approaches, see e.g., in ~\cite{OnlineOpt}--\cite{hu2018distributed}. To minimize content-access latency, energy, storage or bandwidth utilization costs, regularization and decomposition techniques have been used in~\cite{OnlineOpt}. Similar approaches are followed in \cite{Cacherent} to relax a non-convex optimization problem to allocate limited caching memory across a network while accounting for the spatio-temporal content popularity together with the rented storage price fluctuations. An online mixed-integer programming formulation has also been investigated in~\cite{OptOnline}.  

Different from \cite{OnlineOpt,Cacherent,OptOnline}, this paper considers a generic formulation of the problem by introducing time-varying and stochastic costs, and aims at designing more flexible caching schemes, while enabling SBs to learn the optimal fetching-caching decisions. In particular, the fetching and caching decisions are found as the solution of a constrained optimization with the objective of reducing the overall cost, aggregated across files and time instants. Since the caching decision in a given time slot not only affects the instantaneous  cost, but also will influence cache availability in the future, the problem is indeed a dynamic programming (DP). First, by assuming a known stationary distribution for costs as well as popularities, the proposed generic optimization problem is shown to become separable across files, and thus it can be efficiently  solved by decomposing  the so-called value function associated with the original DP into a summation of  smaller-dimension value functions. To reduce the computational complexity, the corresponding marginalized version of the value iteration algorithm \cite{Luismi2014DP_CRs} is introduced, and its performance is assessed via numerical tests. Subsequently, it is shown that having a limited caching capacity and unknown underlying distributions for pertinent parameters, is indeed a special case of this generic formulation. Thus, in order to address caching under limited storage capacity, a dual decomposition technique is developed to cope with the coupling constraint associated with the storage limitation. An online low complexity (marginalized) $Q$-learning based solver is put forth for learning the optimal fetch-cache decisions in an online fashion. The proposed approach is guaranteed to learn optimal fetching-caching decisions in stationary settings, but numerical tests corroborate its improved performance even in non-stationary scenarios.

The rest of this paper is organized as follows. Section~\ref{Sec_Formulation} provides a generic formulation of the problem, where solvers adopted from reinforcement learning are developed in Section~\ref{Sec:DP_Formulation}. Limited storage and back-haul transmission rate settings are discussed in Section~\ref{Sec_limited_storage_and_coms}. Section~\ref{Sec_results} reports numerical results, and finally section~\ref{Sec_Conclusion} provides concluding remarks.

\section{Operating conditions and costs}
\label{Sec_Formulation}
Consider a memory-enabled SB responsible for serving file (content) requests denoted by $f=1,2,\ldots,F$ across time. The requested contents are transmitted to users either by fetching through a (costly)  back-haul transmission link connecting the SB to the cloud, or, by utilizing the local storage unit in the SB where popular contents have been proactively cached ahead of time. The system is considered to operate in a slotted fashion with $t = 1,2, \ldots$ denoting time. 

During slot $t$ and given the available cache contents, the SB receives a number of file requests whose provision incurs certain costs. Specifically, for a requested file $f$, fetching it from the cloud through the back-haul link gives rise to scheduling, routing and transmission costs, whereas its availability at the cache storage in the SB will eliminate such expenses. However, local caching also incurs a number of (instantaneous) costs corresponding to memory or energy consumption. This gives rise to an inherent  caching-versus-fetching trade-off, where one is promoted over the other depending on their relative costs. The objective here is to propose a simple yet sufficiently general framework to minimize the sum-average cost over time by optimizing fetch-cache decisions while adhering to the constraints inherent to the operation of the system at hand, and user-specific requirements. The variables, constraints, and costs involved in  this optimization are described in the ensuing subsections.

\subsection{Variables and constraints}
Consider the system at time slot $t$, where the binary variable $r^f_t$ represents the incoming request for file $f$; that is, $r^f_t = 1$ if  the file $f$ is requested during slot $t$, and $r^f_t = 0$, otherwise. Here, we assume that $r^f_t = 1$ necessitates serving the file to the user and dropping requests is not allowed; thus, requests must be carried out either by fetching the file from the cloud or by utilizing the content currently available in the cache. Furthermore, at the end of each slot, the SB will decide if content $f$ should be stored in the cache for its possible reuse in a subsequent slot.

To formalize this, let us define the ``fetching'' \textit{decision} variable $w^f_t \in \{0,1\}$ along the ``caching'' \textit{decision} variable $a^f_t \in \{0,1\}$. Setting $w^f_t = 1$ implies ``fetching'' file $f$ at time $t$, while $w^f_t=0$ means ``no-fetching.'' Similarly, $a^f_t = 1$ implies that content $f$ will be stored in  cache at the end of slot $t$ for the next slot, while $a^f_t = 0$ implies that it will not. Furthermore, let the storage \textit{state} variable $s_t^f\in\{0,1\}$ account for the availability of files at the local cache. In particular, $s_t^f=1$ if file $f$ is available in the cache at the beginning of slot $t$, and $s_t^f=0$ otherwise. 
Since the availability of file $f$ directly depends on the caching decision at time $t-1$, we have 
\begin{equation}
\label{c1}
{\textrm {C1:}}   \quad s_t^f = a^f_{t-1},   \quad \forall f,t,
\end{equation}
which will be incorporated into our optimization as constraints.

Moreover, since having  $r_t^f=1$ implies transmission of file $f$ to the user(s), it requires either having the file in cache ($s_t^f=1$) or fetching it from the cloud ($w_t^f=1$), giving rise to the second set of constraints 
\begin{equation}
\label{c2}
\textrm {C2:} \quad r^f_t \le w^f_t+s^f_t, \quad  \forall f,t.
\end{equation}  
Finally, the caching decision $a_t^f$ can be set to $1$ only when the content $f$ is available at time $t$; that is, only if either fetching is carried out ($w_t^f=1$) or the current cache state is  $s_t^f=1$. This in turn implies the third set of constraints as    
\begin{equation}
\label{c3}
\textrm {C3:} \quad a^f_t \le s^f_t + w^f_t, \quad  \forall f,t.
\end{equation}

\subsection{Prices and aggregated costs}
To account for the caching and fetching costs, let $\rho_t^f$ and $\lambda_t^f$ denote the (generic) costs associated with $a_t^f=1$ and $w_t^f=1$, respectively. Focusing for now on the caching cost and with $\sigma_f$ denoting the size of content $f$, a simple form for $\rho_t^f$ is 
\begin{equation}\label{eq_generic_form_rho_intro}
\rho_t^f=\sigma_f({\rho'}_t+{\rho'}_t^f) + ({\rho''}_t+{\rho''}_t^f),
\end{equation}
where the first term is proportional to the file size $\sigma_f$, while the second one is constant. Note also that we consider file-dependent costs (via variables ${\rho'}_t^f$ and ${\rho''}_t^f$), as well as cost contributions which are common across files (via ${\rho'}_t$ and ${\rho''}_t$). In most practical setups, the latter will dominate over the former. For example, the caching cost per bit is likely to be the same regardless of the particular type of content, so that ${\rho'}_t^f={\rho''}_t^f=0$. From a modeling perspective, variables $\rho_t^f$ can correspond to actual prices paid to an external entity (e.g., if associated with energy consumption costs), marginal utility or cost functions, congestion indicators, Lagrange multipliers associated with constraints, or linear combinations of those (see, e.g., \cite{Luismi2014DP_CRs,Neely2016ReourceAllocationTutorialBook,Tianyi2017DistributedCloudNets,tps2015wang} and Section \ref{Sec_limited_storage_and_coms}). Accordingly, the corresponding form for the fetching cost is
\begin{equation}\label{eq_generic_form_lambda_intro}
\lambda_t^f=\sigma_f({\lambda'}_t+{\lambda'}_t^f) + ({\lambda''}_t+{\lambda''}_t^f).
\end{equation}
As before, if the transmission link from the cloud to the SB is the same for all contents, the prices ${\lambda'}_t$ and ${\lambda''}_t$ are expected to dominate their file-dependent counterparts ${\lambda'}_t^f$ and ${\lambda''}_t^f$.

\begin{figure} 
	\centering	
	\includegraphics[width=0.37 \textwidth]{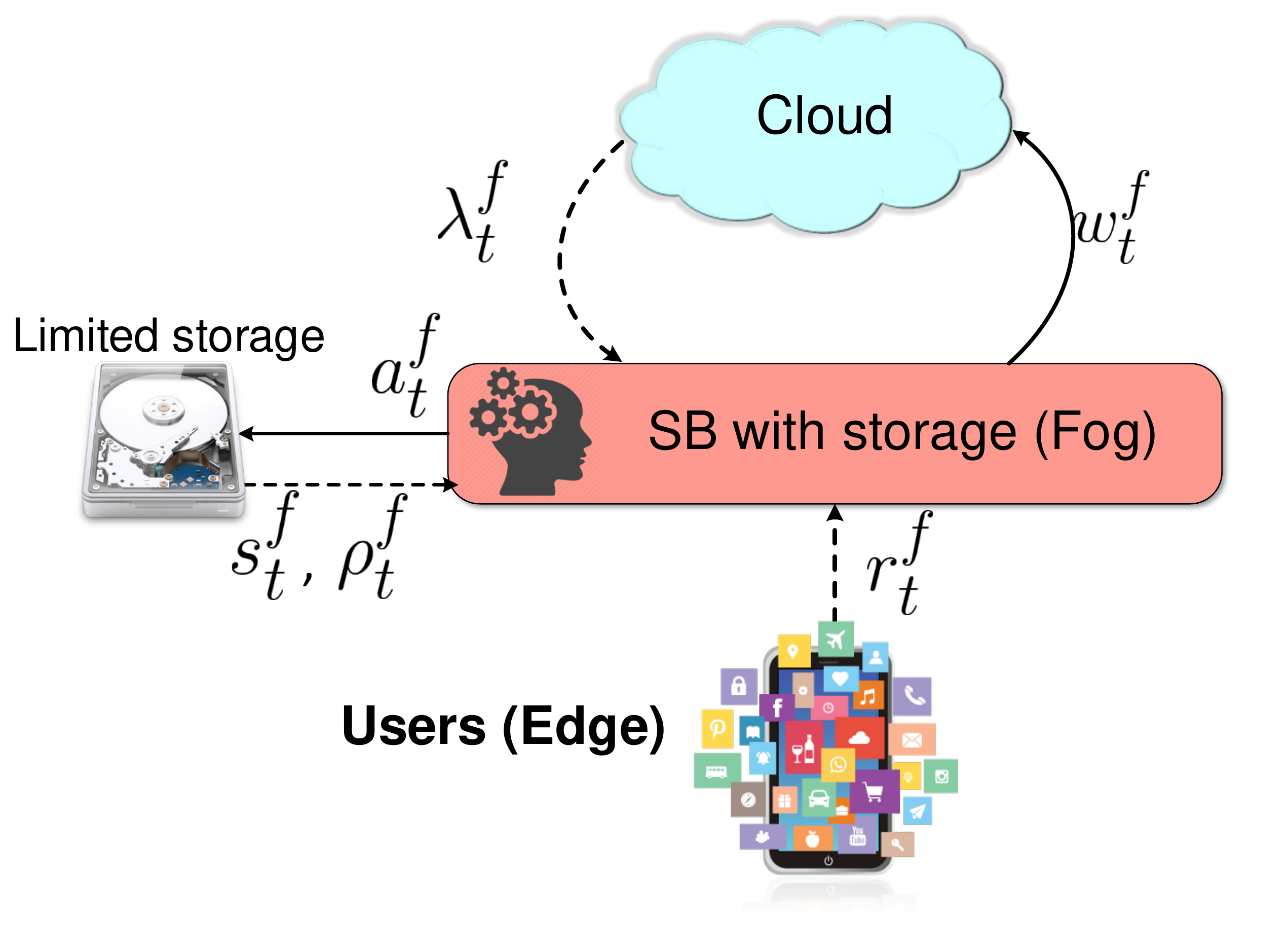}	\caption{System model and main notation. The state variables (dashed lines) are the storage indicator $s_t^f$ and the content request $r_t^f$, as well as the dynamic caching and fetching prices $\rho_t^f$ and $\lambda_t^f$. The optimization variables (solid lines) are the caching and fetching decisions $a_t^f$ and $w_t^f$. The instantaneous per-file cost is $c_t^f=\rho_t^fa_t^f+\lambda_t^fw_t^f$. Per slot $t$, the SB collects the state variables $\{s_t^f,r_t^f;\rho_t^f,\lambda_t^f\}_{f=1}^F$, and decides the values of $\{a_t^f,w_t^f\}_{f=1}^F$ considering not only the cost at time $t$ but also the cost at time instants $t'>t$. }	
	\label{Model} 
\end{figure}

Upon defining the corresponding cost for a given file as $c^f_t (a^f_t,w^f_t;\rho_t^f,\lambda_t^f)=\rho_t^fa^f_t+\lambda_t^f w^f_t $, the aggregate cost at time $t$ is given by 
\begin{equation}
\label{Sum_cost}
	c_t := \sum_{f=1}^F c^f_t (a^f_t,w^f_t;\rho_t^f,\lambda_t^f)=\sum_{f=1}^F \rho_t^fa^f_t + \lambda_t^f w^f_t, 
\end{equation}
which is the basis for the DP formulated in the next section. For future reference, Fig.~\ref{Model} shows a schematic of the system model and the notation introduced in this section. 

\section{Optimal caching with time-varying costs}\label{Sec:DP_Formulation}
Since decisions are coupled across time [cf. constraint \eqref{c1}], and the future values of prices as well as state variables are inherently random, our goal is to optimize the long-term average discounted aggregate cost
\begin{align}
 \label{eq.11}
{\cal {\bar C}} := {\mathbb E} \; \left[\sum _{t=0}^{\infty} \sum  _{f=1}^{F} \gamma ^{t} c^f_t \left(a^f_t,w^f_t;\rho^f_t,\lambda^f_t\right)\right]
 \end{align}
where the expectation is taken with respect to (w.r.t.) the random variables $\boldsymbol \theta_t^f := \{r_t^f,\lambda_t^f,\rho_t^f\}$, and $0<\gamma<1$ is the discounting factor whose tuning trades off current versus more uncertain future costs~\cite[p.44]{sutton1998reinforcement}.  

First, we investigate a setup where the knowledge of the realization of the random variables is causal, that is, the exact value of $\boldsymbol \theta_t^f$ is revealed at the beginning of each slot $t$, and fetch-cache decisions are made sequentially per slot. In addition, the variables in $\boldsymbol \theta_t^f$ are assumed to have stationary and known distributions (e.g., estimated through historical data), which allows for practical estimates of the expectation. Hence, the goal is to take \textit{real-time} fetch-cache decisions by minimizing the expected \textit{current plus future cost} while adhering to operational constraints, giving rise to the following optimization 
\begin{align}
{\textrm {(P$1$)} }  \min  \limits_{ \{(w^f_{k},a^f_{k}) \}_{f,k\geq t}}    &{\cal {\bar C}}_t:=\sum \limits_{k=t}^{\infty} \sum  \limits_{f=1}^{F} \gamma ^{k-t} {\mathbb E}  \left[c^f_k \left(a^f_k,w^f_k;\rho^f_k,\lambda^f_k\right)\right]  \nonumber\\
\mathrm{s.t.}\;\;\;\; &(w^f_{k},a^f_{k}) \in \mathcal{X}( r^f_{k}, a^f_{k-1}),\;\;\;\; \forall f,\,\,k\geq t \nonumber
\end{align}
where 
\begin{align}
\nonumber
\mathcal{X}( r^f_k, a^f_{k-1})&:= \Big\{(w,a)\;\Big|\; w\in\{0,1\}, \,a\in\{0,1\}, \\ \nonumber & s_k^f = a^f_{k-1} , \;  r^f_{k} \le w + s^f_{k},\;\; a \le s^f_{k} + w\Big\},
\end{align}
and the expectation is taken w.r.t.  $\{ \boldsymbol \theta^f_k \}_{\forall k\ge t+1}$.


The presence of constraint \eqref{c1}, which has been made explicit in the definition of $\mathcal{X}( r^f_k, a^f_{k-1})$, implies that current caching decisions impact future costs, and therefore such costs must be taken into account when making the decisions. This ultimately implies that (P$1$) is a DP \cite[p.~79]{sutton1998reinforcement} and, therefore, to solve it we need to: a) identify the current and expected future aggregate cost (this second term will give rise to the so-called value-function); b) write the corresponding Bellman equations; and c) propose a method to estimate the value function. This is the subject of the ensuing subsections, which start by further exploitation of problem structure to reduce complexity.

\subsection{Bellman equations for the per-content problem}
Focusing on (P$1$), one can readily deduce that: (i) consideration of the content-dependent prices renders the objective in (P$1$) separable across $f$, and (ii) the constraints in (P$1$) are also separable across $f$. Furthermore,  the decisions $a_t^f$ and $w_t^f$ for a given $f$, do not affect the values (distribution) of $ \boldsymbol \theta_{k'}^{f'}$ for files $f'\neq f$ and for times $t'>t$. Thus, (P$1$) naturally gives rise to the per-file optimization 
	\begin{align}
	{\textrm {(P$2$)} }  \min  \limits_{ \{(w^f_{k},a^f_{k}) \}_{k\geq t}}    &{\cal {\bar C}}_t^f:=\sum \limits_{k=t}^{\infty}  \gamma ^{k-t} {\mathbb E}  \left[c^f_k \left(a^f_k,w^f_k;\rho^f_k,\lambda^f_k\right)\right]  \nonumber\\
	\mathrm{s.t.}\;\;\;\; &(w^f_{k},a^f_{k}) \in \mathcal{X}( r^f_{k}, a^f_{k-1}),\;\;\;\; k\geq t \nonumber
	\end{align}
which must be solved for $f=1,...,F$. Indeed, the aggregate cost associated with (P$2$) will not depend on variables corresponding to files $f'\neq f$ \cite{Luismi2014DP_CRs}. This is the case if, for instance, the involved variables are independent of each other (which is the setup considered here), or when the focus is on a large system where the contribution of an individual variable to the aggregate network behavior is practically negligible.

\begin{figure*}[t!]
	\setcounter{MYtempeqncnt}{\value{equation}}
	\begin{align} 
	\label{long1} \left(w_t^{f\ast}, a_t^{f\ast}\right) :=& \!\!\!\!\!\underset{(w,a) \in \mathcal{X}( r^f_t, { a^f_{t-1}})}{\arg\mathop{\min}} \left \{ {\mathbb E}_{\boldsymbol \theta_k^f} \left[\underset{(w_k,a_k)\in {\cal X}(r^f_k,{ a^f_{k-1}})}{\min} \left\{\sum \limits_{k=t}^{\infty} \gamma^{k-t} \left[c^f_k(a^f_k,w^f_k;\rho^f_k,\lambda^f_k) \Big | a^f_t \!=\! a, w^f_t \!=\! w, {\boldsymbol \theta^f_t = {\boldsymbol \theta}^f_0} \right]\right\}\right] \right\}\\ 
	 \label{long2}  = &\!\!\!\!\! \underset{(w,a) \in \mathcal{X}( r^f_t, { a^f_{t-1}})}{\arg\mathop{\min}} \left\{c^f_t(a,w;\rho_t^f,\lambda_t^f)  +   {\mathbb {E}}_{\boldsymbol \theta_k^f} \left[ \underset{(w_k,a_k)\in {\cal X}(r^f_k,{ a^f_{k-1}})}{\min}\sum \limits_{k=t+1}^{\infty} \gamma^{k-t}  \left[c^f_k(a^f_k,w^f_k;\rho^f_k,\lambda^f_k) \Big | s^f_{t+1} = a \right]  \right] \right \}
	\end{align} 
\begin{align}
		& \label{value_function}  V^f\left(s^f,r^f;\rho^f,\lambda^f\right) := \!\!\!\!\!\!\!\!\underset{(w,a) \in \mathcal{X}( r^f_t, { a^f_{t-1}})}{\mathop{\min}} \left \{ {\mathbb E}_{\boldsymbol \theta_k^f} \left[\underset{(w_k,a_k)\in {\cal X}(r^f_k,{ a^f_{k-1})}}{\min} \left\{\sum \limits_{k=t}^{\infty} \gamma^{k-t} \left[c^f_k(a^f_k,w^f_k;\rho^f_k,\lambda^f_k) \Big | a^f_t \!=\! a, w^f_t \!=\! w, {\boldsymbol \theta^f_t = {\boldsymbol \theta}^f} \right]\right\}\right] \right\} 
			\end{align} 
			\begin{align}
		 \nonumber 
		 {\bar V}^f(s^f) :=& 
		{\mathbb E}_{\boldsymbol \theta^f} \left[        \underset{(w,a) \in \mathcal{X}( r^f_t, { a^f_{t-1}})}{\mathop{\min}} \left \{ {\mathbb E}_{\boldsymbol \theta_k^f} \left[\underset{(w_k,a_k)\in {\cal X}(r^f_k,{ a^f_{k-1})}}{\min} \left\{\sum \limits_{k=t}^{\infty} \gamma^{k-t} \left[c^f_k(a^f_k,w^f_k;\rho^f_k,\lambda^f_k) \Big | a^f_t \!=\! a, w^f_t \!=\! w, {\boldsymbol \theta^f_t = {\boldsymbol \theta}^f} \right]\right\}\right] \right\}\right] \\
			 \label{bellman} =  &{\mathbb E}_{\boldsymbol \theta^f}  \mathop {\min} \limits_{(w,a) \in \mathcal{X}( r^f, s^f)} \left\{ c^f_0(a,w;\rho^f,\lambda^f)   + \gamma {\bar V^f}(a)  \right\}
		\\ \nonumber
		\end{align} 
	\hrulefill
\end{figure*}

\noindent \textit{Bellman equations and value function:} The DP in (P$2$) can be solved with the corresponding Bellman equations, which require finding the associated value functions \cite[p.~68]{sutton1998reinforcement}. To this end, consider the system at  time $t$, where the cache state as well as the file requests and cost parameters are all given, so that we can write $s^f_t = s^f_{0}$ and $\boldsymbol \theta^f_t = \boldsymbol \theta^f_0$. Then, the optimal fetch-cache decision  $(w^{f \ast}_t,a^{f \ast}_t)$ is readily expressible  as the  solution to \eqref{long1}. The objective in \eqref{long1} is rewritten in \eqref{long2} as the summation of current and discounted average future costs. The form of \eqref{long2} is testament to the fact that problem (P$2$) is a DP and the caching decision $a$ influences not only the current cost $c_t^f(\cdot)$, but also future costs through the second term as well. Bellman equations can be leveraged for tackling such a DP. Under the stationarity assumption for variables $r_{t}^{f}$, $\rho_{t}^{f}$ and $\lambda_{t}^{f}$, the term accounting for the future cost can be rewritten in terms of the \textit{ stationary value function} $V^f\left(s^f,r^f;\rho^f,\lambda^f\right)$ \cite[p.~68]{sutton1998reinforcement}. This function, formally defined in \eqref{value_function}, captures the minimum sum average cost for the ``state'' $(s^f,r^f)$, parametrized by $(\lambda^f,\rho^f)$, where for notational convenience, we define  $\boldsymbol \theta^f := [r^f,\rho^f,\lambda^f]$. 

\subsection{Marginalized value-function}
If one further assumes that price parameters and requests are i.i.d. across time, it can be shown  that the  optimal solution to (P$2$) can be expressed in terms of the  \textit{reduced value function}~\cite{Luismi2014DP_CRs}
\begin{equation}
{\bar V}^f \left(s^f\right) := {\mathbb {E} _ {\boldsymbol \theta^f}} \left[V^f\left(s^f,r^f;\rho^f, \lambda^f\right)\right],
\end{equation}
where the expectation is w.r.t ${\boldsymbol \theta^f}$. This is important not only because it captures the average future cost  of file $f$ for cache state $s^f\in\{0,1\}$, but also because ${\bar V}^f (\cdot)$ is a function of a binary variable,  and  therefore its estimation requires only estimating two values. This is in contrast with the original four-dimensional value function in \eqref{value_function}, whose estimation is more difficult due to its continuous arguments. 

By rewriting the proposed alternative value function ${\bar V}^f (\cdot)$ in a recursive fashion as the summation of instantaneous cost and discounted future values ${\bar V}^f (\cdot)$,  one readily arrives at the Bellman equation form provided in \eqref{bellman}. Thus, the problem reduces to finding $\bar{V}^f(0)$ and $\bar{V}^f(1)$ for all $f$, after which the optimal fetch-cache decisions $(w_t^{f\ast}, a^{f\ast}_t)$ are easily found as the solution to 

  \begin{align}
{\textrm {(P$3$)} }\quad  \min \limits_{(w,a)} \;\;\;\;&  c^f_t(a,w;\rho^f_t,\lambda^f_t)   + \gamma {\bar V}^f \left(a\right)  \nonumber\\
\mathrm{s.t.}\;\;\;\; &(w,a) \in \mathcal{X}( r^f_t, a^f_{t-1}). \nonumber
 \end{align}

If the value-function is known, so that we have access to  ${\bar V}^f(0)$ and ${\bar V}^f(1)$, the corresponding optimal (Bellman) decisions can be found as
\begin{subequations} 
	\begin{align} 
	\!\!&w_t^f=a_t^f,~a_t^f = \mathbb{I}_{\{\Delta {\bar V}^f_\gamma \geq \lambda_t^f + \rho_t^f \}} &  \text{if}~ (r_t^f,s_t^f)=(0,0) \label{E:Bellman_input00}\\
	\!\!&w_t^f=0,~a_t^f = \mathbb{I}_{\{\Delta {\bar V}^f_\gamma \geq \rho_t^f \}} &  \text{if}~ (r_t^f,s_t^f)=(0,1)\\
	\!\!&w_t^f=1,~a_t^f = \mathbb{I}_{\{\Delta {\bar V}^f_\gamma \geq \rho_t^f \}} &  \text{if}~ (r_t^f,s_t^f)=(1,0)\\
	\!\!&w_t^f=0,~a_t^f = \mathbb{I}_{\{\Delta {\bar V}^f_\gamma \geq  \rho_t^f \}}&  \text{if}~ (r_t^f,s_t^f)=(1,1)\label{E:Bellman_input11}
	\end{align}
\end{subequations}
where $\Delta {\bar V}^f_\gamma $ represents the \textit{future} marginal cost, which is obtained as $\Delta {\bar V}^f_\gamma = \gamma (  {\bar V}^f(1) -  {\bar V}^f(0))$, and $\mathbb{I}_{\{ \cdot \}}$ is an indicator function that yields value one if the condition in the argument holds, and zero otherwise. 

The next subsection discusses how ${\bar V}^f(0)$ and ${\bar V}^f(1)$ can be calculated, but first a remark is in order.

\vspace{.15cm}
\noindent {\bf{Remark 1 (Augmented value functions)}}. The value function ${\bar V}^f (s^f)$ can be redefined to account for extra information on $r_t^f$, $\rho_t^f$ or $\lambda_t^f$, if available. For instance, consider the case where  the distribution of $r_t^f$ can be parametrized by $p^f$, which measures content ``popularity'' \cite{breslau1999web}. In such cases,  the value function can incorporate the popularity parameter as an additional input to yield ${\bar V}^f (s^f,p^f)$. Consequently, the optimal decisions will depend not only on the current requests and prices, but also on the (current) popularity $p^f$. This indeed broadens the scope of the proposed approach, as certain types of \textit{non-stationarity} in the distribution of $r_t^f$ can be handled by allowing $p^f$ to (slowly) vary with time.

\subsection{Value function in closed form}\label{Subsec:value_function_stationary}

\noindent 
For notational brevity, we have removed the superscript $f$ in this subsection, and use $\bar V_0$ and $\bar V_1$ in lieu of ${\bar V}(0)$, and ${\bar V}(1)$. Denoting the \textit{long-term} popularity of the content as $p:={\mathbb E} [r_t]$, using the expressions for the optimal actions in  \eqref{E:Bellman_input00}-\eqref{E:Bellman_input11}, and leveraging the independence among $r_t$, $\lambda_t$, and $\rho_t$, the expected cost-to-go function can be readily derived as in \eqref{E:Eq_V1_a}-\eqref{E:Eq_V0_b}. The expectation in \eqref{E:Eq_V1_b_0} is w.r.t. $\rho$, while that in~\eqref{E:Eq_V0_b_0} is w.r.t. both~$\lambda$~and~$\rho$. 
\begin{figure*}[t!]

\begin{eqnarray}
\label{E:Eq_V1_a}   
\bar V_1 & =& (1-p) \Big({\mathbb E} \min_{a \in \left\{0,1\right\}} \left[\gamma \bar V_0 (1-a) + ( \rho +\gamma \bar V_1)a~ \Big|s = 1, r = 0 \right] \Big) \label{E:Eq_V1_b_0}
 + p \Big({\mathbb E} {\min_{a\in \left\{0,1\right\}}} \left[ \gamma \bar V_0 (1-a) + ( \rho +\gamma \bar V_1)a~\Big|s=1, r= 1 \right] \Big) \nonumber \\ \nonumber \\
&=&\gamma \bar V_0 \Pr \big( \rho \geq \Delta \bar V_{\gamma}\big) + {\mathbb E} \Big( \rho  + \gamma \bar V_1 \Big|\rho < \Delta \bar V_{\gamma} \Big)  \Pr \big( \rho < \Delta \bar V_{\gamma}\big)\label{E:Eq_V1_b}
\end{eqnarray}

\begin{eqnarray}
\label{E:Eq_V0_b_0} \bar V_0 &=& (1-p) \Big({\mathbb E} {\min_{a\in\left\{0,1\right\}}} \left[\gamma \bar V_0 (1-a) + (\lambda + \rho +\gamma \bar V_1)a~\Big|s=0, r=0 \right] \Big)  
\\ \nonumber &+& p \Big({\mathbb E} {\min_{a \in \left\{0,1\right\}} } \left[(\lambda + \gamma \bar V_0) (1-a) + (\lambda + \rho +\gamma \bar V_1)a~\Big|s=0,r=1] \right] \Big) \label{E:Eq_V0_b} \\ 
&=& (1-p) \Big( \gamma \bar V_0 \Pr \left(\lambda + \rho \geq \Delta \bar V_{\gamma}\right) +  {\mathbb E} \left( \lambda + \rho +\gamma \bar V_1~\Big|\lambda + \rho < \Delta \bar V_{\gamma}\right)   \Pr \left( \lambda + \rho < \Delta \bar V_{\gamma} \right) \Big)
\\ \nonumber &+& p \Big({\mathbb E} [\lambda ] + \gamma \bar V_0 \Pr \left(\rho \ge \Delta \bar V_\gamma \right) +{\mathbb E} \left(\rho + \gamma \bar V_1~\Big|\rho \le \Delta \bar V_\gamma \right) \Pr \left(\rho \le \Delta \bar V_\gamma \right) \Big)
\end{eqnarray}
\hrulefill
\end{figure*}

Solving the system of equations in \eqref{E:Eq_V1_a}-\eqref{E:Eq_V0_b} yields the optimal values for $\bar V_1$ and $\bar V_0$. A simple solver would be to perform exhaustive search over the range of these values since it is only a two-dimensional search space.  However, a better alternative to solving the given system of equations is to rely on the well known \textit{value iteration} algorithm \cite[p.~100]{sutton1998reinforcement}. In short, this is an offline algorithm, which per iteration $i$ updates the  estimates $\{\bar{V}^{i+1}_0 , \bar{V}^{i+1}_1\}$ by computing the expected cost using $\{\bar{V}^{i}_0 , \bar{V}^{i}_1\}$, until the desired accuracy is achieved. This scheme is tabulated in detail in Algorithm~1, for which the distributions of $r, \rho, \lambda$ are assumed to be known. We refer to \cite[p.100]{sutton1998reinforcement} for a detailed discussion on the value-iteration algorithm, and its convergence guarantees. 

\begin{algorithm}[t]
	\SetKwInOut{Input}{Input}
	\SetKwInOut{Output}{Output}
	\underline{Set} $\bar V^0_0 = \bar V^0_1 = 0$ \;
	
	\Input{$\gamma<1$, probability density function of $\rho,\lambda$ and $r$,  precision $\epsilon$, in order to stop}
	\Output{$\bar V_0$, $\bar V_1$}
	\While {$|\bar V^i_s-\bar V^{i+1}_s| < \epsilon; s \in \left\{0,1\right\}$} { \For{$s = 0,1$} {$\bar{V}^{i+1}_s = \mathbb{E}_{r,\rho,\lambda}\mathop {\min} \limits_{(w,a) \in \mathcal{X}( r, s)} \left\{  c(a,w;\rho,\lambda) + \gamma {\bar V^i_a}\right\}$}
		$i = i+1$}    
	\caption{Value iteration  for finding $\bar V \left(\cdot\right)$}
\end{algorithm}

\vspace{.15cm}
\noindent \textbf{Remark 2 (Finite-horizon approximate policies).} In the proposed algorithms, namely exhaustive search as well as Algorithm~1, the solver is required to compute an expectation, which can be burdensome in setups with limited computational resources. For such scenarios,  the class of finite-horizon policies emerges as a computationally affordable suboptimal alternative \cite[p. 242]{sutton1998reinforcement}. The idea behind such policies is to truncate the infinite summation in the objective of (P$1$); thus, only considering the impact of the current decision on a few number of future time instants denoted by $h$, typically referred to as the \textit{horizon}. The extreme case of a finite-horizon policy is that of a \textit{myopic policy} with $h=0$, which ignores any future impact of current decision,  a.k.a. zero-horizon policy, thus taking the action which minimizes the instantaneous cost. This is equivalent to setting the future marginal cost to zero, hence solving \eqref{E:Bellman_input00}-\eqref{E:Bellman_input11} with  $\Delta {\bar V}_\gamma=\Delta {\bar V}_\gamma^{h=0} = 0$. 

Another commonly used alternative is to consider the impact of the current decision for only the next time instant, which corresponds to the so-called horizon-1 policy. This  entails setting the future cost at $h=1$ as $\Delta {\bar V}_\gamma^{h=1} = \gamma (\bar V_1^{h=0} - \bar V_0^{h=0})$ with
\begin{eqnarray}
\!\! \bar V_0^{h=0} \!\!& = & \!\!\! (1-p) {\mathbb E}[\lambda w^{h=0}+\rho a^{h=0}|s=0,r=0] \nonumber \\
\!\!\!\!& + &\!\!\! p {\mathbb E}[\lambda w^{h=0}\!+\!\rho a^{h=0}|s=0,r=1]=p{\mathbb E}[\lambda] \label{eq_value_function_finite_horizon0}\\
\!\!\bar V_1^{h=0} \!\!& = & \!\!\!(1-p) {\mathbb E}[\lambda w^{h=0}+\rho a^{h=0}|s=1,r=0] \nonumber \\ 
\!\!\!\!&+& \!\!\!p {\mathbb E}[\lambda w^{h=0}+\rho a^{h=0}|s=1,r=1]=0,\label{eq_value_function_finite_horizon1}
\end{eqnarray}
which are then substituted into \eqref{E:Bellman_input00}-\eqref{E:Bellman_input11} to yield the actions $ w^{h=1}$ and  $a^{h=1}$. The notation $ w^{h=0}$ and  $a^{h=0}$ in \eqref{eq_value_function_finite_horizon0} and \eqref{eq_value_function_finite_horizon1} is used to denote the actions obtained when \eqref{E:Bellman_input00}-\eqref{E:Bellman_input11} are solved using the future marginal cost at horizon zero $\Delta {\bar V}_\gamma^{h=0}$, which as already mentioned, is zero; that is, under the myopic policy in lieu of the original optimal solution. Following an inductive argument, the future marginal cost at $h=2$ is obtained as $\Delta {\bar V}_\gamma^{h=2} = \gamma (\bar V_1^{h=1} - \bar V_0^{h=1})$ with
\begin{eqnarray}
	\nonumber
	\!\! \bar V_0^{h=1} \!\!& = & \!\!\! (1-p) {\mathbb E}[\lambda w^{h=1}+\rho a^{h=1}+ \gamma \bar{V}_a^{h=0}|s=0,r=0] \\\nonumber 
	\!\!\!\!& + &\!\!\! p {\mathbb E}[\lambda w^{h=1}\!+\!\rho a^{h=1}+ \gamma \bar{V}_a^{h=0}|s=0,r=1], \\ 	\nonumber
	\!\!\bar V_1^{h=1} \!\!& = & \!\!\!(1-p) {\mathbb E}[\lambda w^{h=1}+\rho a^{h=1}+\gamma \bar{V}_a^{h=0}|s=1,r=0] \\  \nonumber
	\!\!\!\!&+& \!\!\!p {\mathbb E}[\lambda w^{h=1}+\rho a^{h=1}+\gamma \bar{V}_a^{h=0}|s=1,r=1],
\end{eqnarray}
which will allow to obtain the actions $ w^{h=2}$ and  $a^{h=2}$. While increasing horizons can be used, as $h$ grows large, solving the associated equations becomes more difficult and computation of the optimal stationary policies, is preferable. 

\subsection{State-action value function ($Q$-function):}\label{Subsec:q_function_stationary}
In many practical scenarios, knowing the underlying distributions for $\rho_t$, $\lambda_t$ and $r_t$ may not be possible, which motivates the introduction of online solvers that can learn the parameters on-the-fly. As clarified in the ensuing sections, in such scenarios, the so-called $Q$-function (or state-action value function)~\cite[p.69]{sutton1998reinforcement} becomes helpful, since there are rigorous theoretical guarantees on the convergence of its stochastic estimates; see~\cite{tsitsiklis} and \cite{watkins1992q}. 
Motivated by this fact, instead of formulating our dynamic program using the value (cost-to-go) function, we can alternatively formulate it using the $Q$-function. 
Aiming at an online solver, let us tackle the DP through the estimation (learning) of the $Q$-function. Equation\footnote{Equations \eqref{Q_function}-\eqref{marg_q}, and \eqref{eq_Q_factors_esemble} are shown at the top of page 7.} \eqref{Q_function} defines the $Q$-function for a specific file under a given state $\left(s_t,r_t\right)$, parametrized by cost parameters $\left(\rho_t,\lambda_t\right)$. Under stationarity distribution assumption for $\left\{ \rho_t, \lambda_t,r_t \right\}$, the $Q$-function $ Q\left(s_t,r_t, w_t,a_t;\rho_t,\lambda_t\right)$ accounts for the minimum average aggregate cost at state $\left(s_t,r_t\right)$, and taking specific fetch-cache decision $(w_t,a_t)$ as for the first decision, while followed by the best possible decisions in next slots. This function is parametrized by $\left(\rho_t,\lambda_t\right)$ since while making  the current cache-fetch decision, the current values for these cost parameters are assumed to be known. The original $Q$-function in \eqref{Q_function} needs to be learned over all values of $\left\{s_t,r_t, w_t,a_t,\rho_t, \lambda_t, r_t\right\}$, thus suffering from the curse of dimensionality, especially due to the fact that $\rho_t$ and $\lambda_t$ are continuous variables. 

\begin{figure*}[t!]
	\setcounter{MYtempeqncnt}{\value{equation}}
	\begin{align} 
	\label{Q_function}  Q\left(s_t,r_t, {w_t,a_t};\rho_t,\lambda_t\right) & := {\mathbb E} \left[\underset{\left\{(w_k,a_k)\in {\cal X}(r_k,{ a_{k-1}})\right\}_{k=t+1}^{\infty}}{\min} \left\{\sum \limits_{k=t}^{\infty} \gamma^{k-t} \left[c_k(a_k,w_k;\rho_k,\lambda_k) \Big |a_t, w_t, {\boldsymbol \theta_t = {\boldsymbol \theta}} \right]\right\}\right] \\   & \hspace{-2cm} = \underbrace{c_{ t}(a_t,w_t;\rho_t,\lambda_t)}_{\textrm{Immediate cost}} + \gamma \underbrace{{\mathbb E} \left[\underset{\left\{(w_k,a_k)\in {\cal X}(r_k,{ a_{k-1}})\right\}_{k=t+1}^{\infty}}{\min} \left\{\sum \limits_{k=t+1}^{\infty} \gamma^{k-{(t+1)}} \left[c_k(a_k,w_k;\rho_k,\lambda_k) \Big | s_{t+1} = a_t \right]\right\}\right]}_{\textrm{Average minimum future cost}}
	\end{align} 
	\begin{align}
	\label{marg_q}
	{\bar Q}_{r_t,s_t}^{w_t,a_t}:= &\; {{\mathbb E}}_{ \rho_t, \lambda_t} \; \left[ Q\left(s_t,r_t, { w_t,a_t };\rho_t,\lambda_t\right) \right], \quad \forall (w_t,a_t) \in { \mathcal X}(r_t,{{a_{t-1}}}) \\ = &\; {\mathbb E}_{ \rho_t, \lambda_t} \left[c_{ t}(a_t,w_t;\rho_t,\lambda_t) \right]  + \gamma \left[ {\mathbb E}_{{\boldsymbol \theta}_{t+1}} \left[ Q\left(s_{t+1},r_{t+1}, { w_{t+1}^{\ast},a_{t+1}^{\ast}};\rho_{t+1},\lambda_{t+1}\right) \Big| {{\boldsymbol \theta}_{t+1}}, s_{t+1} = a_t \right] \right]. 
	\nonumber
\end{align}
\setcounter{equation}{\value{MYtempeqncnt}} 
\hrulefill
\setcounter{equation}{\value{MYtempeqncnt}+4} 
\begin{align}\label{eq_Q_factors_esemble}
	\bar Q_{r,s}^{w,a} =  {\mathbb E} [\lambda] w +  {\mathbb E} [\rho] a +  \gamma (1-p) \hspace{-.75 cm}  \sum_{\forall (z_1,z_2){ \in \mathcal{X}(0,a)}} \hspace{-0.5 cm} \bar Q_{0,a}^{z_1,z_2} \Pr \Big((w_{t+1}^*,a_{t+1}^*)=(z_1,z_2)|(s_{t+1},r_{t+1})=(a,0)\Big)\nonumber && \\   +\gamma p  \sum_{\forall (z_1,z_2) { \in \mathcal{X}(1,a)}} \bar Q_{1,a}^{z_1,z_2} \Pr \Big((w_{t+1}^*,a_{t+1}^*)=(z_1,z_2)|(s_{t+1},r_{t+1})=(a,1)\Big). && \end{align}
\setcounter{equation}{\value{MYtempeqncnt}+3}
	\hrulefill
\end{figure*}
To alleviate this burden, we define the \textit{marginalized} $Q$-function $ Q(s_t,r_t, w_t,a_t)$ in \eqref{marg_q}. By changing the notation for clarity of exposition, the marginalized $Q$-function, $\bar Q_{r_t,s_t}^{w_t,a_t}$, can be rewritten in a more compact form as 
\begin{equation}
\label{M_Q}
\bar Q_{r_t,s_t}^{w_t,a_t} = {\mathbb E} \Big[\lambda_t w_t + \rho_t a_t + \gamma  \bar Q_{r_{t+1},a_t}^{w_{t+1}^*,a_{t+1}^*} \Big] {\; \forall (w_t,\!a_t) \!\in\! { \mathcal {X}}(r_t,{a_{t-1}})}.
\end{equation}
Note that, while the marginalized value-function is only a function of the state, the marginalized $Q$-function depends on both the state $\left(r,s\right)$ and the immediate action $\left(w,a\right)$. The main reason one prefers to learn the value-function rather than the $Q$-function is that the latter is computationally more complex. To see this, note that the input space of $\bar Q_{r_t,s_t}^{w_t,a_t}$ is a four-dimensional binary space, hence the function has $2^4=16$ different inputs and one must estimate the corresponding $16$ outputs. Each of these possible values are called $Q$-factors, and under the stationarity assumption, they can be found using \eqref{eq_Q_factors_esemble} 
defined for all $(r,s,w,a)$. In this expression, we have $(z_1,z_2)\in\{0,1\}^2$ and the term $\Pr \left( \left(w_{t+1}^\ast a^{\ast}_{t+1}\right) = (z_1,z_2)\right)$ stands for the probability of specific action $(z_1, z_2)$ to be optimal at slot $t+1$. This action is random because the optimal decision at $t+1$ depends on $\rho_{t+1}$, $\lambda_{t+1}$ and $r_{t+1}$, which are not known at slot $t$.  Although not critical for the discussion, if needed, one can show that half of the 16 $Q$-factors can be discarded, either for being infeasible -- recall that $(w_t,a_t) \!\in\! { \mathcal {X}}(r_t,{a_{t-1}})$ -- or suboptimal. This means that \eqref{eq_Q_factors_esemble} needs to be computed only for $8$ of the $Q$-factors.

From the point of view of offline estimation, working with the $Q$-function is more challenging than working with the $V$-function, since more parameters need to be estimated. In several realistic scenarios however, the distributions of the state variables are unknown, and one has to resort to stochastic schemes in order to learn the parameters on-the-fly. In such scenarios, the $Q$-function based approach  is preferable, because it enables learning the optimal decisions in an online fashion even when the underlying distributions are unknown. 

\subsection{Stochastic policies: Reinforcement learning}\label{Subsec:value_q_function_stochastic}
As discussed in Section \ref{Subsec:value_function_stationary}, there are scenarios where obtaining the optimal value function (and, hence, the optimal stationary policy associated with it) is not computationally feasible. The closing remark in that section discussed policies which, upon replacing the optimal value function with approximations easier to compute, trade reduced complexity for loss in optimality. However, such reduced-complexity methods still require knowledge of the state distribution [cf. \eqref{eq_value_function_finite_horizon0} and \eqref{eq_value_function_finite_horizon1}]. In this section, we discuss stochastic schemes to approximate the value function under unknown distributions. The policies resulting from such stochastic methods offer a number of advantages since they: (a) incur a reduced complexity; (b) do not require knowledge of the underlying state distribution; (c) are able to handle some non-stationary environments; and in some cases, (d) they come with asymptotic optimality guarantees. To introduce this scheme, we first start by considering a simple method that updates stochastic estimates of the value function itself, and then proceed to a more advanced method which tracks the value of the $Q$-function. Specifically, the presented method is an instance of the celebrated $Q$-learning algorithm \cite{watkins1989learning}, which is the workhorse of stochastic approximation in DP \cite[p.~68]{sutton1998reinforcement}.

\vspace{.1cm}
\subsubsection{Stochastic value function estimates}
The first method relies on current stochastic estimates of $\bar V_0$ and $\bar V_1$, denoted by $\hat{\bar V}_0(t)$ and $\hat{\bar V}_1(t)$ at time $t$ (to be defined rigorously later). Given $\hat{\bar V}_0(t)$ and $\hat{\bar V}_1(t)$ at time $t$, the (stochastic) actions $\hat{w}_t$ and $\hat{a}_t$  are taken via solving \eqref{E:Bellman_input00}-\eqref{E:Bellman_input11} with $\Delta {\bar V}_\gamma=\gamma(\hat{\bar V}_0(t)-\hat{\bar V}_1(t))$. Then, stochastic estimates of the value functions $\hat{\bar V}_0(t)$ and $\hat{\bar V}_1(t)$ are updated as
\begin{itemize}
	\item If $s_t=0$, then $\hat{\bar V}_1(t+1)=\hat{\bar V}_1(t)$ and 
	$\hat{\bar V}_0(t+1)= (1-\beta) \hat{\bar V}_0(t) + \beta (\hat{w}_t\lambda_t + \hat{a}_t\rho_t  + \gamma \hat{\bar V}_{\hat{a}_t}(t) )$;
	\item If $s_t=1$, then $ \hat{\bar V}_0({t+1})=\hat{\bar V}_0({t})$ and 
	$\hat{V}_1(t+1)= (1-\beta) \hat{\bar V}_1(t) + \beta (\hat{w}_t\lambda_t + \hat{a}_t\rho_t  + \gamma \hat{\bar V}_{\hat{a}_t}(t) ) $;
\end{itemize}
where $\beta >0$ denotes the stepsize. While easy to implement (only two recursions are required), this algorithm has no optimality guarantees.

\vspace{.1cm}
\subsubsection{\textit{Q}-learning algorithm} Alternatively, one can run a stochastic approximation algorithm on the $Q$-function. This entails replacing the $Q$-factors $\bar Q_{r,s}^{w,a}$ with stochastic estimates $\hat{\bar Q}_{r,s}^{w,a}(t)$. To describe the algorithm, suppose for now that at time $t$, the estimates $\hat{\bar Q}_{r,s}^{w,a}(t)$ are known for all $(r,s,w,a)$. 
Then, in a given slot $t$ with $(r_{t}, s_{t})$,  action $ \left(\hat{w}^\ast_t,\hat{a}^\ast_t\right)$ is obtained via either an exploration or an exploitation step. When exploring, which happens with a small probability $\epsilon_t$, a random and feasible action $ \left(\hat{w}^\ast_t,\hat{a}^\ast_t\right) \in {\mathcal X} \left(r_t,a_{t-1}\right)$ is taken. In contrast, in the exploitation mode, which happens with a probability $1 - \epsilon_t$,  the optimal action according to the current estimate of $\hat{\bar Q}_{r,s}^{w,a}(t)$ is 
\begin{equation} 
\setcounter{equation}{\value{equation}+1}
\left(\hat{w}^\ast_t,\hat{a}^\ast_t\right) := \underset{(w,a) \in \mathcal{X}( r_{t}, a_{t-1})}{\arg \min} \;  w\lambda_t + a \rho_t + \gamma \hat{\bar Q}_{r_t,s_t}^{w,a}(t) 
\label{margQ_solver}.
\end{equation} 
After taking this action, going to next slot $t+1$, and observing $\rho_{t+1}, \lambda_{t+1}$, and $r_{t+1}$, the $Q$-function estimate is updated as
\begin{flalign}
\label{Q_update}
&\hat{\bar Q}_{r,s}^{w,a}(t+1) = \nonumber \\
&\begin{cases}
\!\hat{\bar Q}_{r,s}^{w,a}(t) \quad {\text{ if} }\quad  (r,s,w,a)\neq(r_t,s_t,\hat{w}^\ast_t,\hat{a}^\ast_t) \\ \\
\!\!(1\!-\!\beta) \hat{\bar Q}_{r_t,s_t}^{\hat w^{\ast}_t,\hat a^{\ast}_t}(t)  +  \beta \Big(\hat{w}_t^{\ast} \lambda_t + \hat{a}_t^{\ast} \rho_t  + {\gamma \hat{\bar Q}_{r_{t+1},\hat a_t^\ast}^{\hat w_{t+1}^{\ast}, \hat a_{t+1}^{\ast}}(t)\Big)} \;\text{o.w.,} 
\end{cases}
\end{flalign}
where ``o.w.'' stands for ``otherwise,'' and $(\hat w^{\ast}_{t+1},\hat a^{\ast}_{t+1})$ is the optimal action for the next slot. 
This update rule describes one of the possible implementations of the $Q$-learning algorithm, which was originally introduced in \cite{watkins1989learning}. This online algorithm enables making sequential decisions in an unknown environment, and is guaranteed to learn optimal decision-making rules under certain conditions \cite[p.148]{sutton1998reinforcement}, 
\cite{watkins1992q}.  The aforementioned exploration-exploitation step is necessary for the factors  $\hat{\bar Q}_{r,s}^{w,a}$ to converge to their optimal value ${{\bar Q}^{\ast w,a}}_{\;\;r,s}$  \cite{ODE}, \cite{tsitsiklis}. Intuitively, under continuous updates of \textit{all} state-action pairs  along with regular stochastic approximation conditions on the stepsize $\beta$, the updates on $\hat{\bar Q}_{r,s}^{w,a}$ converge to the optimal values  with probability 1. Various exploration-exploitation algorithms have been proposed to meet convergence guarantees~\cite[p.~839]{russell2016artificial} . A necessary condition for any such exploration-exploitation approach is the \textit{greedy in the limit of infinite exploration} (GLIE) property~\cite[p.~840]{russell2016artificial}. A common choice to meet this property is the $\epsilon$-greedy approach with $\epsilon_t = 1/t$, providing guaranteed yet slow convergence. In practice however, $\epsilon_t$ is set to a small value for faster convergence; see ~\cite{ODE} and \cite{tsitsiklis} for a more detailed discussion on convergence.  

The resultant algorithm for the problem at hand is tabulated in Algorithm~\ref{Q_learning}. It is important to stress that in our particular case, we expect the algorithm to converge fast. That is the case because, under the decomposition approach followed in this paper as well as the introduction of the marginalized $Q$-function, the state-action space of the resultant $Q$-function has very low dimension and hence, only a small number of $Q$-factors need to be estimated.  

\begin{algorithm}[t]
	\SetKwInOut{Input}{Input}
	\SetKwInOut{Output}{Output}
	\Input{$0< \gamma, \beta < 1$}
	\Output{$ \hat {\bar Q}_{r,s}^{w,a}(t+1)$}
{\textbf{Initialize} $\hat {\bar Q}_{r,s}^{w,a}(1) = 0$, $s_1 = 0$, {$\{r_0, \rho_0, \lambda_0\}$ are revealed}	\\
	\For {$t = 1 , 2, \ldots$}{
	{\textrm	For the current state $(r_t,s_t)$, choose $(\hat w_t^{\ast},\hat a_t^{\ast})$ } {\[(\hat {w}_t^{\ast},\hat {a}_t^{\ast})  = \left\{
			\begin{array}{ll}
			{\textrm {Solve \eqref{margQ_solver}}} &  \textrm{w.p. } 1-\epsilon_t \\
			\textrm{random } (w ,a ) \in {\mathcal X}_t(r_t,s_t)  & \textrm{w.p. }  \epsilon_t
			\end{array}
			\right. \] \\} {Update state $s_{t+1} = \hat a_t^{\ast}$} \\ 
		{Request and cost parameters, $ \boldsymbol \theta_{t+1}$, are revealed} \\ 
		{Update $Q$ factor by \eqref{Q_update}}}
	\caption{$Q$-learning algorithm to estimate ${\bar Q}_{r,s}^{w,a}$ for a given file $f$}
	\label{Q_learning}}
\end{algorithm}

\section{Limited storage and back-haul transmission rate via dynamic pricing}\
\label{Sec_limited_storage_and_coms}

So far, we have considered that the prices $\{\rho_t^f,\lambda_t^f\}$ are provided by the system, and we have not assumed any explicit limits (bounds) neither on the capacity of the local storage nor on the back-haul transmission link between the SB and the cloud. 	
In this section, we discuss such limitations, and describe how by leveraging dual decomposition techniques, one can redefine the prices $\{\rho_t^f,\lambda_t^f\}$ to account for capacity constraints. 

\subsection{Limiting the instantaneous storage rate}

In this subsection, practical limitations on the cache storage capacity are explored. Suppose that the SB is equipped with a \textit{single} memory device that can store $M$ files. Clearly, the cache decisions should then satisfy the following constraint per time slot 	
\begin{equation}\nonumber
{\textrm {C$4$:}}\quad \sum \limits_{f = 1}^{F} a_t^f \sigma^f \le M, \quad t = 1,2, \ldots 
\label{eq20}
\end{equation}	
In order to respect such hard capacity limits, the original optimization problem in (P$1$) can be simply augmented with C$4$, giving rise to a new optimization problem which we will refer to as (P$4$).  
Solving (P$4$) is more challenging than (P$1$), since the constraints in C$4$ must be enforced at each time instant, which subsequently couples the optimization across files. In order to deal with this, one can dualize C$4$ by augmenting the cost with the primal-dual term $\mu_t(\sum_{f=1}^F\sigma_fa_t^f-M)$, where $\mu_t$ denotes the Lagrange multiplier associated with the capacity constraint C$4$. The resultant problem is separable across files, but requires finding $\mu_t^*$, the optimal value of the Lagrange multiplier, at each and every time instant. 

If the solution to the original unconstrained problem (P$1$) does satisfy C$4$, then  $\mu_t^* = 0 $ due to complementary slackness. On the other hand, if the storage limit is violated, then the constraint is active, the Lagrange multiplier satisfies $\mu_t^*>0$, and its exact value must be found using an iterative algorithm. Once the value of the multiplier is known, the optimal actions associated with (P$4$) can be found using the expressions for the optimal solution to (P$1$) provided that the original storage price $\rho_t^f$ is replaced with the new storage price $\rho_{t,aug}^f=\rho_t^f+\mu_t^*\sigma_f$ {[cf. \eqref{eq_generic_form_rho_intro}]}. The reason for this will be explained in detail in the following subsection, after introducing the ensemble counterpart of C$4$.  

\subsection{Limiting the long-term storage rate}\label{subsec_long-term_storage_rate}
Consider now the following constraint [cf. C$4$] 
\begin{equation}
	\textrm{C$5$:} \quad  \sum \limits_{k=t}^{\infty} \gamma^{k-t} {\mathbb {E}}\left[  \sum \limits_{f = 1}^{F} a_k^f \sigma^f  \right]  \le \sum \limits_{k=t}^{\infty} \gamma^{k-t} M'
\label{relaxedconstraint}
\end{equation}
where the expectation is taken w.r.t.  all state variables. 
By setting $M'=M$, one can view C$5$ as a relaxed version of C$4$. That is, while C$4$ enforces the limit to be respected at every time instant, C$5$ only requires it to be respected \textit{on average}. From a computational perspective, dealing with C$5$ is easier than  its instantaneous counterpart, since in the former only one constraint is enforced and, hence, only one Lagrange multiplier, denoted by $\mu$, must be found. This comes at the  price that guaranteeing C$5$ with $M'=M$ does not imply that C$4$ will always be satisfied. Alternatively, enforcing C$5$ with $M'<M$, will increase the probability of satisfying C$4$, since the solution will guarantee that ``on average'' there exists free space on the cache memory. A more formal discussion on this issue will be provided in the remark closing the subsection.

To describe in detail how accounting for C$5$ changes the optimal schemes, let (P$5$) be the problem obtained after augmenting (P$1$) with C$5$. Suppose now that to solve (P$5$) we dualize the single constraint in C$5$. Rearranging terms, the augmented objective associated with (P$5$) is given by
\begin{align}\label{eq_aumented_objective_storage_multiplier}
\sum \limits_{k=t}^{\infty} \sum  \limits_{f=1}^{F} \gamma ^{k-t} {\mathbb E}  
\left[c^f_k \left(a^f_k,w^f_k;\rho^f_k,\lambda^f_k\right)+ \mu a_k^f \sigma^f \right]  
- \sum \limits_{k = t}^{\infty} \gamma^{k-t}  M'.
\end{align}
Equation \eqref{eq_aumented_objective_storage_multiplier} demonstrates that after dualization and provided that the multiplier $\mu$ is known, decisions can be optimized separately across files. To be more precise, note that the term $\sum_{k = t}^{\infty} \gamma^{k-t}  M'$ in the objective is constant, so that it can be ignored, and define the modified instantaneous cost as
\begin{align}
\nonumber 
{\check{c}_k^f} := & \; c^f_k \left(a^f_k,w^f_k;\rho^f_k,\lambda^f_k\right)  + \mu \sigma^f a_k^f \\
= & \; \left(\rho_k^f + \mu \sigma^f \right) a^f_k+\lambda_k^f w^f_k.\label{eq_instantaneous_cost_augmented_storage_mult}
\end{align}
%
%
The last equation not only reflects that the dualization indeed facilitates separate per-file optimization, but it also reveals that term $\mu \sigma^f$ can be interpreted as an additional storage cost associated with the long-term caching constraint. More importantly, by defining the modified (augmented) prices ${\rho^f_{t, \textrm {aug}}} := \rho_t^f + \mu \sigma^f$ for all $t$ and $f$, the optimization of  \eqref{eq_instantaneous_cost_augmented_storage_mult} can be carried out with the schemes presented in the previous sections, provided that $\rho_t^f$ is replaced with ${\rho^f_{t, \textrm {aug}}}$. 

Note however that in order to run the optimal allocation algorithm, the value of $\mu$ needs to be known. Since the dual problem is always convex, one option is to use an iterative dual subgradient method,  which computes the satisfaction/violation of the constraint C$5$ per iteration~\cite{palomar2006tutorial}, \cite[p.223]{boydconvex}. Clearly, this requires knowledge of the state distribution, since the constraint involves an expectation. When such knowledge is not available, or when the computational complexity to carry out the expectations cannot be afforded, stochastic schemes are worth considering. For the particular case of estimating Lagrange multipliers associated with long-term constraints, a simple but powerful alternative is to resort to \textit{stochastic dual} subgradient schemes \cite{palomar2006tutorial}, \cite{boydconvex},  which for the problem at hand, estimate the value of the multiplier $\mu$ at every time instant $t$ using the update rule
\begin{align}
\hat{\mu}_{t+1}  =& \left[ \hat{\mu}_t + \zeta  \left( \sum \limits_{f = 1}^{F} \hat{a}^{f\ast}_t \sigma^f - M' \right) \right]^{+}.
\label{dual_cache}
\end{align}
In the last expression, $\zeta>0$ is a (small) positive constant, the update multiplied by $\zeta$ corresponds to the violation of the constraint after removing the expectation, the notation $[\cdot]^+$ stands for the $\max\{0,\cdot\}$, and $\hat{a}^{f\ast}_t$ denotes the optimal caching actions obtained with the policies described in Section \ref{Sec:DP_Formulation} provided that $\rho_t^f$ is replaced by $\hat{\rho}^f_{t, \textrm {aug}} = \rho_t^f + \hat{\mu}_t \sigma^f$. 

We next introduce another long-term constraint that can be considered to limit the storage rate. This constraint is useful not only because it gives rise to alternative novel caching-fetching schemes, but also because it will allow us to establish connections with well-known algorithms in the area of congestion control and queue management. To start, define the variables $\alpha_{in,t}^f:=[a_t^f-s_t^f]^+$ and $\alpha_{out,t}^f:=[s_t^f-a_t^f]^+$ for all $f$ and $t$. Clearly, if $\alpha_{in,t}^f=1$, then content $f$ that was not in the local cache at time $t-1$, has been stored at time $t$; and as a result, less storage space is available. On the other hand, if $\alpha_{out,t}^f=1$, then content $f$ was removed from the cache at time $t$, thus freeing up new storage space. With this notation at hand, we can consider the long term constraint
\begin{equation}
\textrm{C$6$:} \,  \sum \limits_{k=t}^{\infty} \gamma^{k-t} {\mathbb {E}}\left[  \sum \limits_{f = 1}^{F} \alpha_{in,k}^f \sigma^f  \right]  \le \sum \limits_{k=t}^{\infty} \gamma^{k-t} {\mathbb {E}}\left[  \sum \limits_{f = 1}^{F} \alpha_{out,k}^f \sigma^f  \right],
\label{relaxedconstraint}
\end{equation}
which basically ensures the long-term stability of the local-storage. That is, the amount of data stored in the local memory is no larger than that taken out from the memory, guaranteeing that in the long term stored data does not grow unbounded. 

To deal with C$6$ we can follow an approach similar to that of C$5$, under which we first dualize C$6$ and then use a stochastic dual method to estimate the associated dual variable. With a slight abuse of notation, supposing that the Lagrange multiplier associated with stability is by also denoted $\mu$, the counterpart of \eqref{dual_cache} for the constraint C$6$ is
\begin{align}
\hat{\mu}_{t+1}  = \left[ \hat{\mu}_t + \zeta \sum \limits_{f = 1}^{F}   [\hat{a}^{f\ast}_t - s^f_t]^+  - [s^f_t - \hat{a}^{f\ast}_t]^+ \right]^{+}.
\label{dualupdate_long_term_storage_as_queue}
\end{align}
Note that the update term in the last iteration follows after removing the expectations in C$6$ and replacing $\alpha_{in,t}^f$, and $\alpha_{out,t}^f$ with their corresponding definitions. The modifications that the expressions for the optimal policies require to account for this constraint are a bit more intricate. If $s_t^f=0$, the problem structure is similar to that of the previous constraints, and we just need to replace $\rho_t^f$ with $\hat{\rho}^f_{t, \textrm {aug}} = \rho_t^f + \hat{\mu}_t \sigma^f$. However,  if $s_t^f=1$, it turns out that: i) deciding ${\hat a}_t^{ f \ast}=1$ does not require modifying the caching price, but ii) deciding ${\hat a}_t^{f \ast}=0$ requires considering the \textit{negative} caching price $-\hat{\mu}_t \sigma^f$. In other words, while our formulation in Section \ref{Sec:DP_Formulation} only considers incurring a cost when $a_t^f=1$ (and assumes that the instantaneous cost is zero for $a_t^f=0$), to fully account for C$6$, we would need to modify our original formulation so that costs can be associated with the decision $a_t^f=0$ as well. This can be done either by considering a new cost term or, simply by replacing $\gamma \bar{V}^f(0)$ by $\gamma \bar{V}^f(0) -\hat{\mu}_t \sigma^f$ in \eqref{E:Bellman_input00}-\eqref{E:Bellman_input11}, which are Bellman's equations describing the optimal policies.
 
\vspace{.15cm} 
\noindent \textbf{Remark 3 (Role of the stochastic multipliers).} It is well-established that the Lagrange multipliers can be interpreted as the marginal price that the system must pay to \linebreak (over-)satisfy the constraint they are associated with \cite[p.241]{boydconvex}. When using stochastic methods for estimating the multipliers, further insights on the role of the multipliers can be obtained \cite{Neely2016ReourceAllocationTutorialBook,Marques_Queues12,Tianyi2017DistributedCloudNets}. Consider for example the update in \eqref{dual_cache}. The associated constraint C$5$ establishes that the long-term storage rate cannot exceed $M'$. To guarantee so, the stochastic scheme updates the estimated price in a way that, if the constraint for time $t$ is oversatisfied, the price goes down, while if the constraint is violated, the price goes up. Intuitively, if the price estimate $\hat{\mu}_t$ is far from its optimal value and the constraint is violated for several consecutive time instants, the price will keep increasing, and eventually will take a value sufficiently high so that storage decisions are penalized/avoided. How quickly the system reacts to this violation can be controlled via the constant $\zeta$. Interestingly, by tuning the values of $M'$ and $\zeta$, and assuming some regularity properties on the distribution of the state variables, conditions under which deterministic short-term limits as those in C$4$ are satisfied can be rigorously derived; see, e.g., \cite{Tianyi2017DistributedCloudNets} for a related problem in the context of distributed cloud networks. A similar analysis can be carried out for the update in \eqref{dualupdate_long_term_storage_as_queue} and its associated constraint C$6$. Every time the instantaneous version of the constraint is violated because the amount of data stored in the memory exceeds the amount exiting the memory, the corresponding price $\hat{\mu}_t$ increases, thus rendering future storage decisions more costly. In fact, if we initialize the multiplier at $\hat{\mu}_t=0$ and set $\zeta=1$, then the corresponding price is the total amount of information stored at time $t$ in the local memory. In other words, the update in \eqref{dualupdate_long_term_storage_as_queue} exemplifies how the dynamic prices considered in this paper can be used to account for the actual state of the caching storage. Clearly, additional mappings from the instantaneous storage level to the instantaneous storage price can be considered. The connections between stochastic Lagrange multipliers and storing devices have been thoroughly explored in the context of demand response, queuing management and congestion control. We refer the interested readers to, e.g., \cite{Neely2016ReourceAllocationTutorialBook,Marques_Queues12}.    

\subsection{Limits on the  back-haul transmission rate}\label{subsec_long-term_comm_rate}
The previous two subsections dealt with limited caching storage, and how some of those limitations could be accounted for by modifying the caching price $\rho_t^f$. This section addresses limitations on the back-haul transmission rate between the SB and the cloud as well as their impact on the fetching price $\lambda_t^f$.
 
While our focus has been on optimizing the decisions at the SB, contemporary networks must be designed following a holistic (cross-layer) approach that accounts for the impact of local decisions on the rest of the network. Decomposition techniques (including those presented in this paper) are essential to that end \cite{palomar2006tutorial}. For the system at hand, suppose that $\mathbf{x}_{CD}$ includes all variables at the cloud network, $\bar{C}_{CD}(\mathbf{x}_{CD})$ denotes the associated cost, and the feasible set $\mathcal{X}_{CD}$ accounts for the constraints that cloud variables $\mathbf{x}_{CD}$ must satisfy. Similarly, let $\mathbf{x}_{SB}$, $\bar{C}_{SB}(\mathbf{x}_{SB})$, and $\mathcal{X}_{SB}$ denote the corresponding counterparts for the SB optimization analyzed in this paper. Clearly, the fetching actions $w_t^f$ are included in $\mathbf{x}_{SB}$, while the variable $b_t$ representing back-haul transmission rate (capacity) of the connecting link between the cloud and the SB, is included in $\mathbf{x}_{CD}$. This transmission rate will depend on the resources that the cloud chooses to allocate to that particular link, and will control the communication rate (and hence the cost of fetching requests) between the SB and the cloud. As in the previous section, one could consider two types of capacity constraints
\begin{subequations}
\begin{flalign}\label{eq_constraint_com_limits_short_term}
{\rm C}7a: \quad	&\sum_{f=1}^{F} w^f_t \sigma^f \le b_t, \quad  t = 1, \ldots, \\
{\rm C}7b: \quad \label{eq_constraint_com_limits_long_term}
	 &\sum_{k=t}^\infty \gamma^{k-t} \sum_{f=1}^{F} {\mathbb E} [w^f_t \sigma^f] \le \sum_{k=t}^\infty \gamma^{k-t} {\mathbb E} [b_k],
\end{flalign}
\end{subequations}
depending on whether the limit is imposed in the short term or in the long term.	
	
With these notational conventions, one could then consider the \textit{joint} resource allocation problem
\begin{align}
 \min  \limits_{ \mathbf{x}_{CD}, \mathbf{x}_{SB}}    & \bar{C}_{CD}(\mathbf{x}_{CD}) + \bar{C}_{SB}(\mathbf{x}_{SB})  \nonumber\\
\mathrm{s.t.}\;\;\;\; &\mathbf{x}_{CD}\in \mathcal{X}_{CD},\;\;\mathbf{x}_{SB}\in \mathcal{X}_{SB},\;\;({\rm C}7)
\end{align}
where the constraint C$7$ -- either the instantaneous one in C$7a$ or the lon-term version in C$7b$ -- couples both optimizations. It is then clear that if one dualizes  C$7$, and the value of the Lagrange multiplier associated with C$7$ is known, then two separate optimizations can be run: one focusing on the cloud network and the other one on the SB. For this second optimization, consider for simplicity that the average constraint in \eqref{eq_constraint_com_limits_long_term} is selected and let $\nu$ denote the Lagrange multiplier associated with such a constraint. The optimization corresponding to the SB is then      
\begin{align}
\min  \limits_{ \mathbf{x}_{SB}}    \; \bar{C}_{SB}(\mathbf{x}_{SB}) + \sum_{k=t}^\infty \gamma^{k-t} \sum_{f=1}^{F} {\mathbb E} [w^f_t \nu \sigma^f]\;\;\;\mathrm{s.t.}\;\; \mathbf{x}_{SB}\in \mathcal{X}_{SB}.
\end{align}

Clearly, solving this problem is equivalent to solving the original problem in Section \ref{Sec:DP_Formulation}, provided that the original cost is augmented with the primal-dual term associated with the coupling constraint. To address the modified optimization, we will follow steps similar to those in Section \ref{subsec_long-term_storage_rate}, defining first a stochastic estimate of the Lagrange multiplier as 
\begin{align}
\label{dual_backhaul}
\hat{\nu}_{t+1}  = \left[ \hat{\nu}_t + \zeta  \left( \sum \limits_{f = 1}^{F} \hat{w}^{f\ast}_t \sigma^f - b_t \right) \right]^{+},
\end{align}
and then obtaining the optimal caching-fetching decisions running the schemes in Section \ref{Sec:DP_Formulation} after replacing the original fetching cost $\lambda_t^f$ with the augmented one $\lambda^f_{t,{\textrm {aug}}}=\lambda_t^f+\hat{\nu}_t\sigma_f$.

For simplicity, in this section we will limit our discussion to the case where $\hat{\nu}_t$ corresponds to the value of a Lagrange multiplier corresponding to a communication constraint. However, from a more general point of view, $\hat{\nu}_t$ represents the marginal price that the cloud network has to pay to transmit the information requested by the SB. In that sense, there exists a broad range of options to set the value of $\hat{\nu}_t$, including the congestion level at the cloud network (which is also represented by a Lagrange multiplier), or the rate (power) cost associated with the back-haul link. While a detailed discussion on those options is of interest, it goes beyond the scope of the present work. 

\subsection{Modified online solver based on $Q$-learning}
\label{Qlearning_solver}
We close this section by providing an online reinforcement-learning algorithm that modifies the one introduced in Section \ref{Sec:DP_Formulation} to account for the multipliers introduced in Section \ref{Sec_limited_storage_and_coms}.  
\begin{algorithm}[t]
	\SetKwInOut{Input}{Input}
	\SetKwInOut{Output}{Output}
	\Input{$0 < \gamma, \; \beta < 1, \; \hat \mu_0, \zeta, \epsilon_t, \; M$}  
	\Output{ $\hat {\bar Q}_{r^f,s^f}^{w^f,a^f} \left(t+1\right)$ }
	{\textbf{Initialize} Set $ \hat {\bar Q}_{r^f,s^f}^{w^f,a^f} (1) = 0$ for all factors  
		\linebreak Set $s_0^f = 0$ and variables ${\boldsymbol \theta}^f_{0}=\{r^f_0, \rho^f_0, \lambda^f_0\}$ are revealed}  \\	
	\For {$t = 0,1 \ldots$} 
	{\textrm{For the current state $(r^f_t,s^f_t)$, choose $ ({\breve w}_t^{f \ast},{\breve a}_t^{f \ast})$ }
		 { \[ ({\breve w}_t^{f \ast},{\breve a}_t^{f \ast}) \! = \! \left\{
			\begin{array}{ll}
			{\textrm {Solve \eqref{margQ_solver}}} &  \textrm{w.p. } 1-\epsilon_t \\
			\textrm{random }  (w,a) \! \in \! { \mathcal X}_t^f(r^f_t,s^f_t)  & \textrm{w.p. }  \epsilon_t
			\end{array}
			\right. \] \\}
		{Update dual variable \[\hat \mu_{t+1} = \left[\hat \mu_{t} + \zeta \left( \sum \limits_{f = 1}^{F} {{ \breve{a}}}_t^{f \ast} \sigma^f - M \right) \right]^{+}\] \\ }
				{Incur cost \quad $\check{c}^f_t := c^f_t ({ {\breve a}_t^{f \ast}},{\breve w}_t^{f \ast};\rho^f_t,\lambda^f_t) + \hat \mu_t {\breve  a}^{f\ast}_t \sigma^f$ \\}
		{\textrm (If required) Apply  $\Pi_{\rm C4}(\cdot)$ to guarantee C$4$ {\hspace{+2.5 cm} \[ \Pi_{\rm C4} \left[ \left\{  ({\breve w}_t^{f \ast},{\breve a}_t^{f \ast}) \right\}_f \right] \rightarrow \left\{{w}^{f\ast}_t,{ {a}^{f\ast}_t}\right\}_f \] }   \\}
		{Update state  $s^f_{t+1} = { {a}^{f\ast}_t}$  \\}
		{Request and cost parameters,  $ \boldsymbol \theta^f_{t+1}$, are revealed} \\ 
		{Update all ${\hat {\bar Q}}$  factors as  \\} \nonumber \[\hat {\bar Q}_{r_{t}^f,s^f_{t}}^{w^{f \ast}_{t},a^{f \ast}_{t}}(t+1) = (1-\beta)  \; \hat {\bar Q}_{r_{t}^f,s^f_{t}}^{w^{f \ast}_{t},a^{f \ast}_{t}}(t) \; + \] \[ \hspace{2 cm} \beta   \left[ \check{c}_{t}^f + \nonumber \gamma  \underset{{ (w^f,a^f)} \in { \mathcal X}^f_{t+1} } {\min}  \hat {\bar Q}^{{w^f,a^f}}_{r^f_{t+1},s^f_{t+1}}(t) \right] \]   
	}
	\caption{Modified $Q$-learning for online caching}
	\label{MQ_learning}
\end{algorithm}
By defining per file cost $\hat c_k^f$ as
\begin{align}
\nonumber \hat c_k^f \left(w_k^f,a_k^f;\rho_k^f,\lambda_k^f,{ \hat \mu_k},{ \hat \nu_k}\right) := &  \\ \left(\rho_k^f + \hat \mu_k \sigma^f \right) & a^f_k  + \left( \lambda_k^f + \hat \nu_k \sigma^f\right) w^f_k
\end{align}
the problem of caching under limited cache capacity and back-haul link reduces to per file optimization as follows 
\begin{align}
	\nonumber ({\textrm P}8) \min \limits_{ \{(w^f_{k},a^f_{k}) \}_{k\geq t}}  & \; \sum \limits_{k=t}^{\infty} \gamma ^{k-t} {\mathbb E}  \left[ {\hat{c}_k^f}\left(a^f_k,w^f_k;\rho^f_k,\lambda^f_k,{ \hat \mu_k},{ \hat \nu_k}\right) \right]   \nonumber \\ \nonumber \mathrm{s.t.}\;\;\;\; &(w^f_{k},a^f_{k}) \in \mathcal{X}( r^f_{k}, { a^f_{k-1}}),\;\;\;\; \forall f,\,\,k\geq t
\end{align}
where the updated dual variables $\hat \mu_k$ and $\hat \nu_k$ are obtained respectively by iteration \eqref{dual_cache} and \eqref{dual_backhaul}. If we plug $\hat{c}_k^f$ instead of $c_k^f$ into the marginalized $Q$-function in \eqref{marg_q}, then the solution for (P8) in current iteration $k$ for a given file $f$ can readily be found by solving 
\begin{equation} 
\underset{(w,a) \in \mathcal{X}( r_{t}, {  a_{t-1}})}{\arg \min} \; {\bar Q}_{r_t,s_t}^{w,a} + w (\lambda_t +\hat \nu_t \sigma^f) + a (\rho_t+\hat \mu_t \sigma^f).
\label{light_wieght_solver}
\end{equation}
Thus, it suffices to form a marginalized $Q$-function for each file and solve \eqref{light_wieght_solver}, which can be easily accomplished through exhaustive search over $8$ possible cache-fetch decisions~$ (w,a) \in \mathcal{X} (r_{t}, {  a_{t-1}})$. 

To simplify notation and exposition, we focus on the \textit{limited caching capacity} constraint, and suppose that the back-haul is capable of serving any requests, thus $\hat \nu_t = 0, \; \forall t $. Modifications to account also for  $\hat \nu_t \neq  0$ are straightforward. 

The modified $Q$-learning (MQ-learning) algorithm, tabulated in Algorithm~\ref{MQ_learning}, essentially learns to make optimal fetch-cache decisions while accounting for the limited caching capacity constraint in C$4$ and/or C$5$. In particular, to provide a computationally efficient solver the stochastic updates corresponding to C$5$ are used. Subsequently, if C$4$ needs to be enforced, the obtained solution is projected into the feasible set through projection algorithm $\Pi_{\rm C4}(\cdot)$.  The projection $\Pi_{\rm C4}(.)$ takes the obtained solution $ \{{  \breve w ^{f\ast}_t, {\breve a}^{f\ast}_t } \}_{\forall f}$, the file sizes, as well as the marginalized $Q$-functions as input, and generates a feasible solution $\{{w}^{f\ast}_t,{a}^{f\ast}_t \}_{\forall f}$ satisfying C$4$ as follows: it sorts the files with $ {\breve a}^{f\ast}_t = 1$ in ascending $Q$-function order, and caches the files with the lowest $Q$-values until the cache capacity is reached. Overall, our modified algorithm performs a ``double'' learning: i) by using reinforcement schemes it learns the optimal policies that map states to actions, and ii) by using a stochastic dual approach it learns the mechanism that adapt the prices to the saturation and congestion conditions in the cache. Given the operating conditions and the design approach considered in the paper, the proposed algorithm has moderate complexity, and thanks to the reduced input dimensionality, it also converges in a moderate number of iterations.


%
\section{Numerical tests}
\label{Sec_results}
\begin{figure}
	\centering
	\includegraphics[width=0.48 \textwidth]{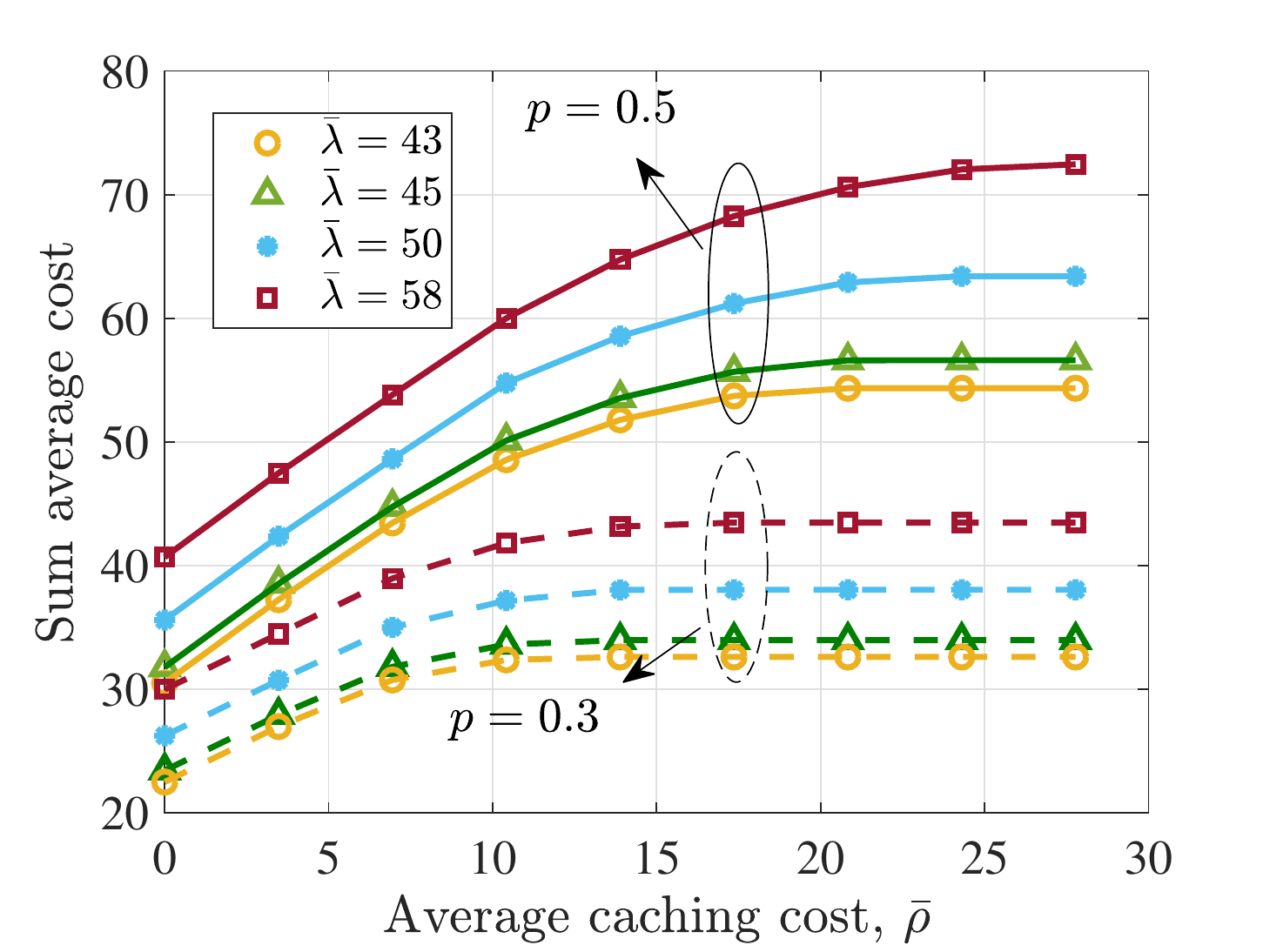}
	\caption{Average cost versus $\bar \rho$ for different values of $p, \bar \lambda$.}
	\label{fig:result1}
\end{figure}

\begin{figure}
	\centering
	\includegraphics[width=0.48 \textwidth]{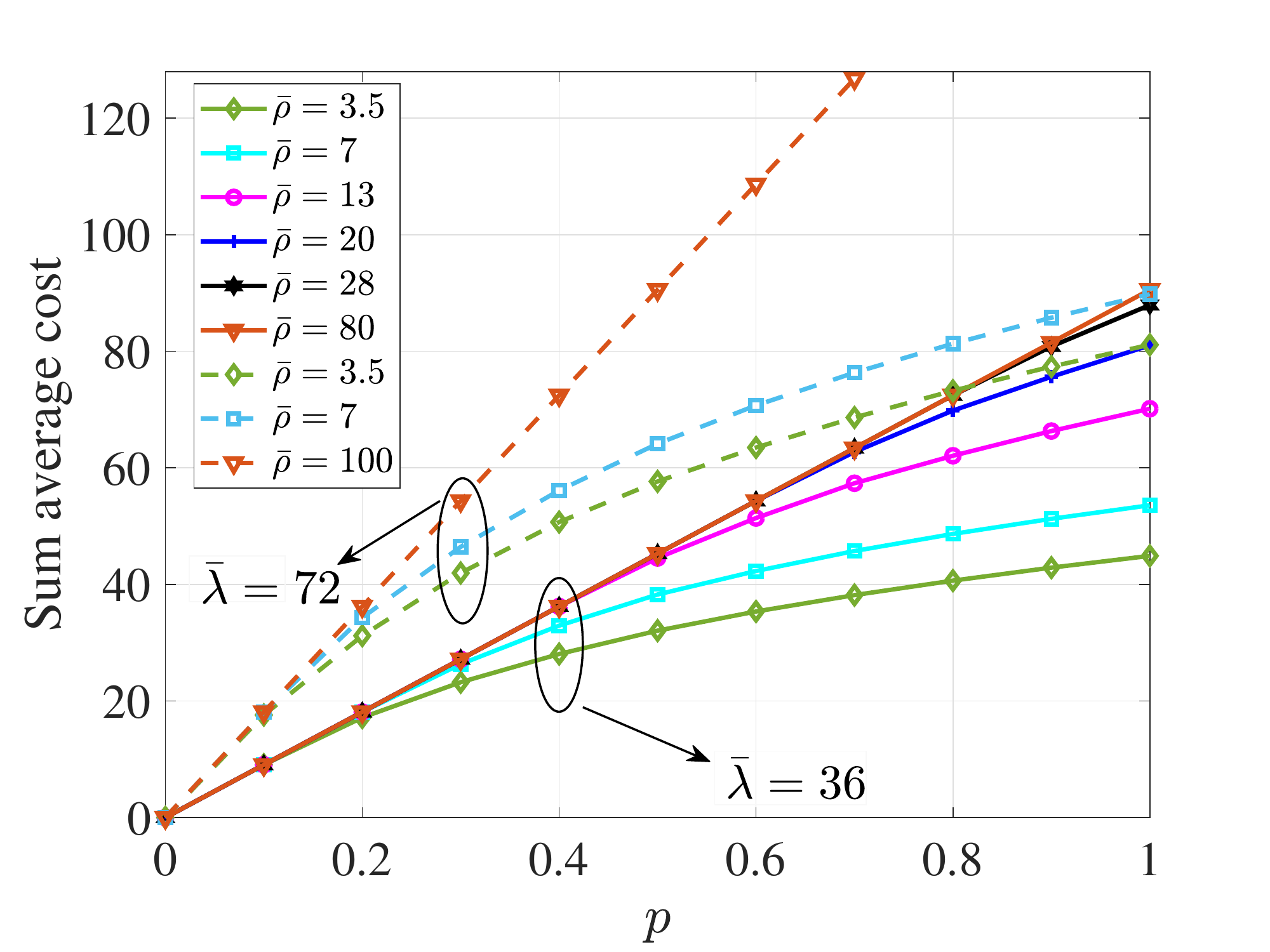}
	\caption{Average cost versus $p$ for different values of $\bar \lambda, \bar \rho$.}
	\label{fig:result2}
\end{figure}

\begin{figure}
	\centering
	\includegraphics[width=0.48 \textwidth]{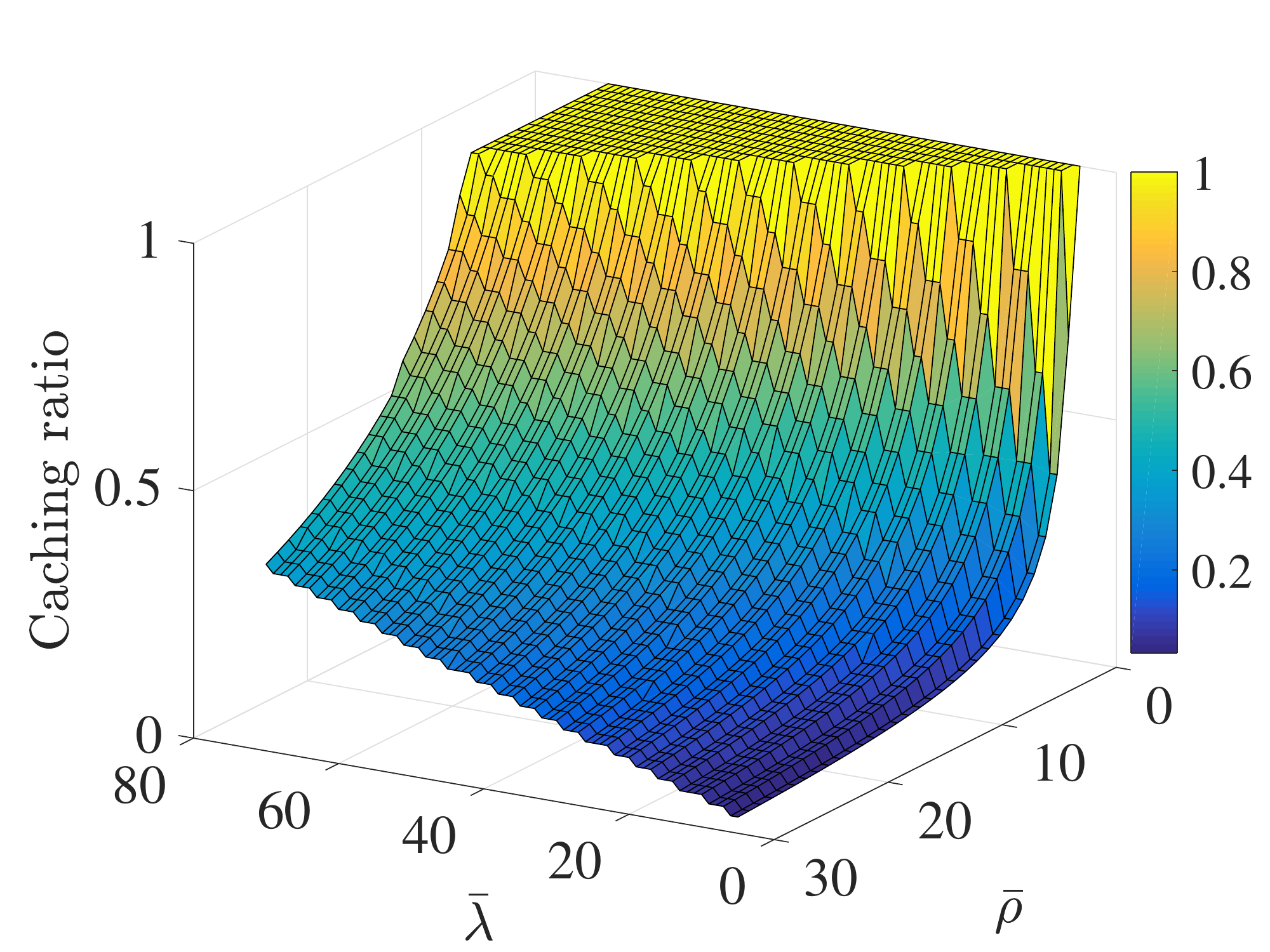}
	\caption{Caching ratio vs. $\bar \rho$ and $\bar \lambda$ for $p = 0.5$ and $s = r = 1$.}
	\label{fig:result3}
\end{figure}

\begin{figure}
	\centering
	\includegraphics[width=0.48 \textwidth]{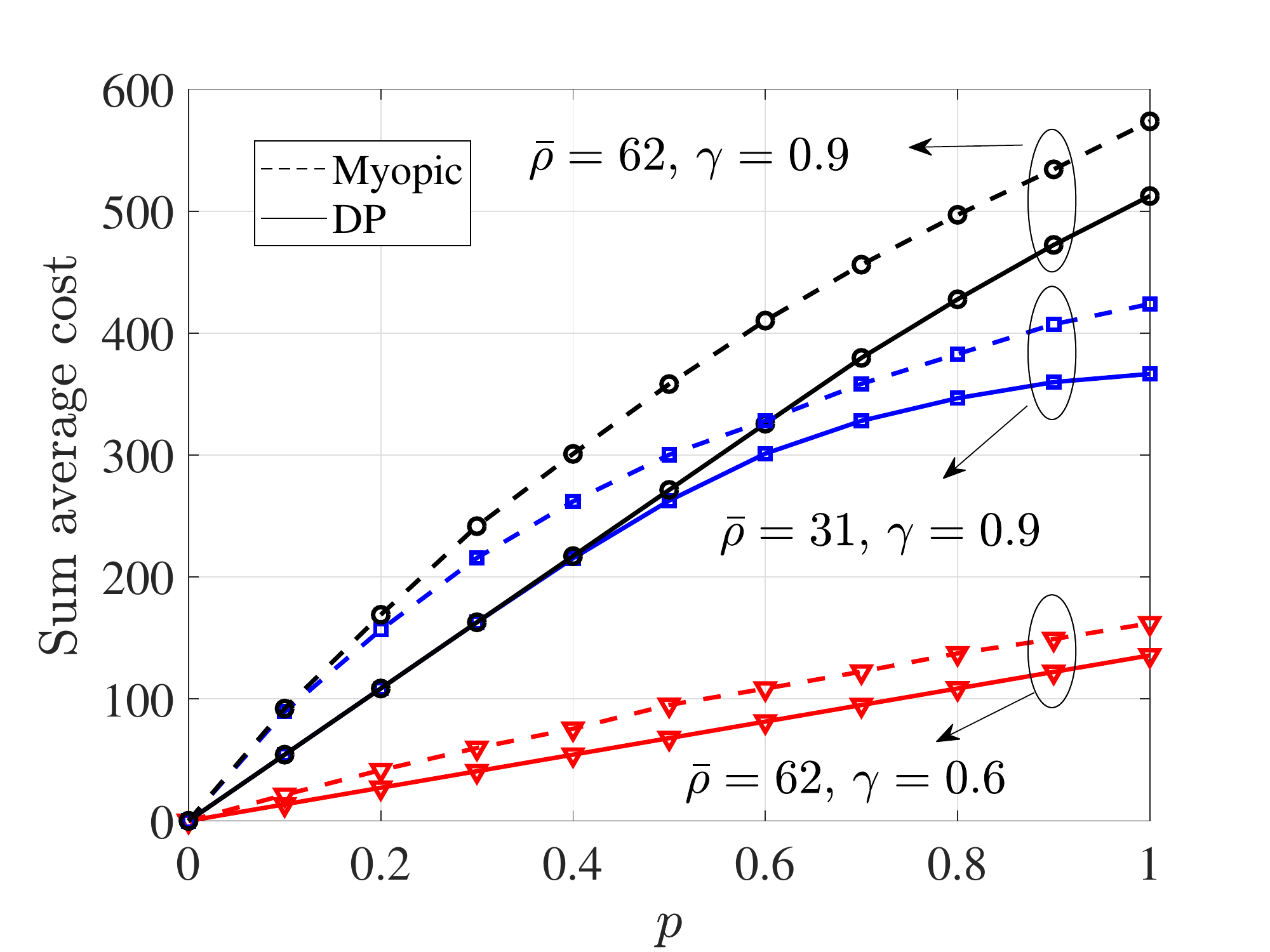}
	\caption{Performance of DP versus myopic caching for $\bar \lambda =53$. }
	\label{fig:result4}
\end{figure}

\begin{figure}
	\centering
	\includegraphics[width=0.5 \textwidth]{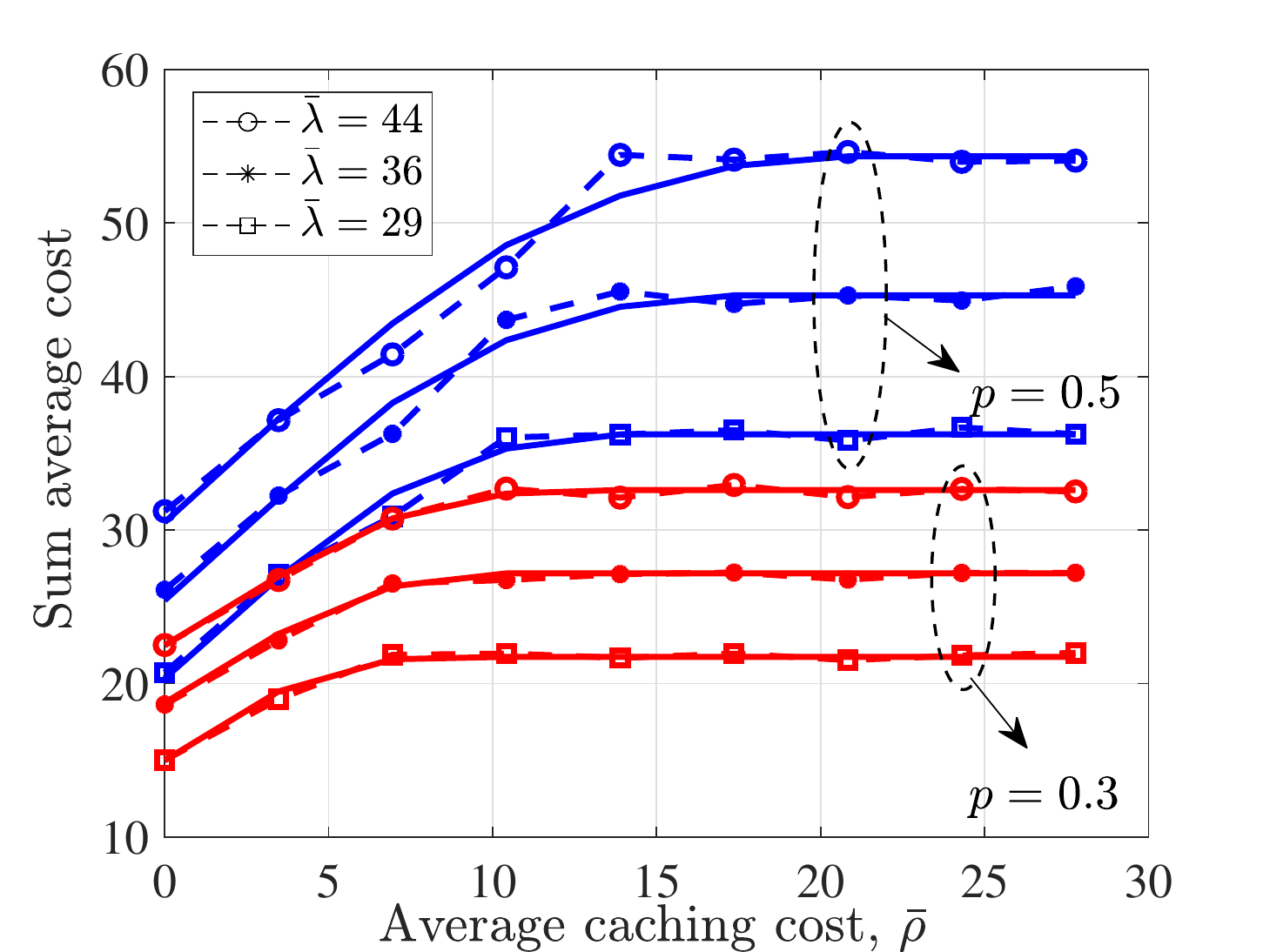}
	\caption{Average cost versus ${\bar \rho}$ for different values of $\bar \lambda, p$. Solid line is for value iteration while dashed lines are for $Q$-learning based solver.}
	\label{result6}
\end{figure}

\begin{figure}
	\centering
	\includegraphics[width=0.5 \textwidth]{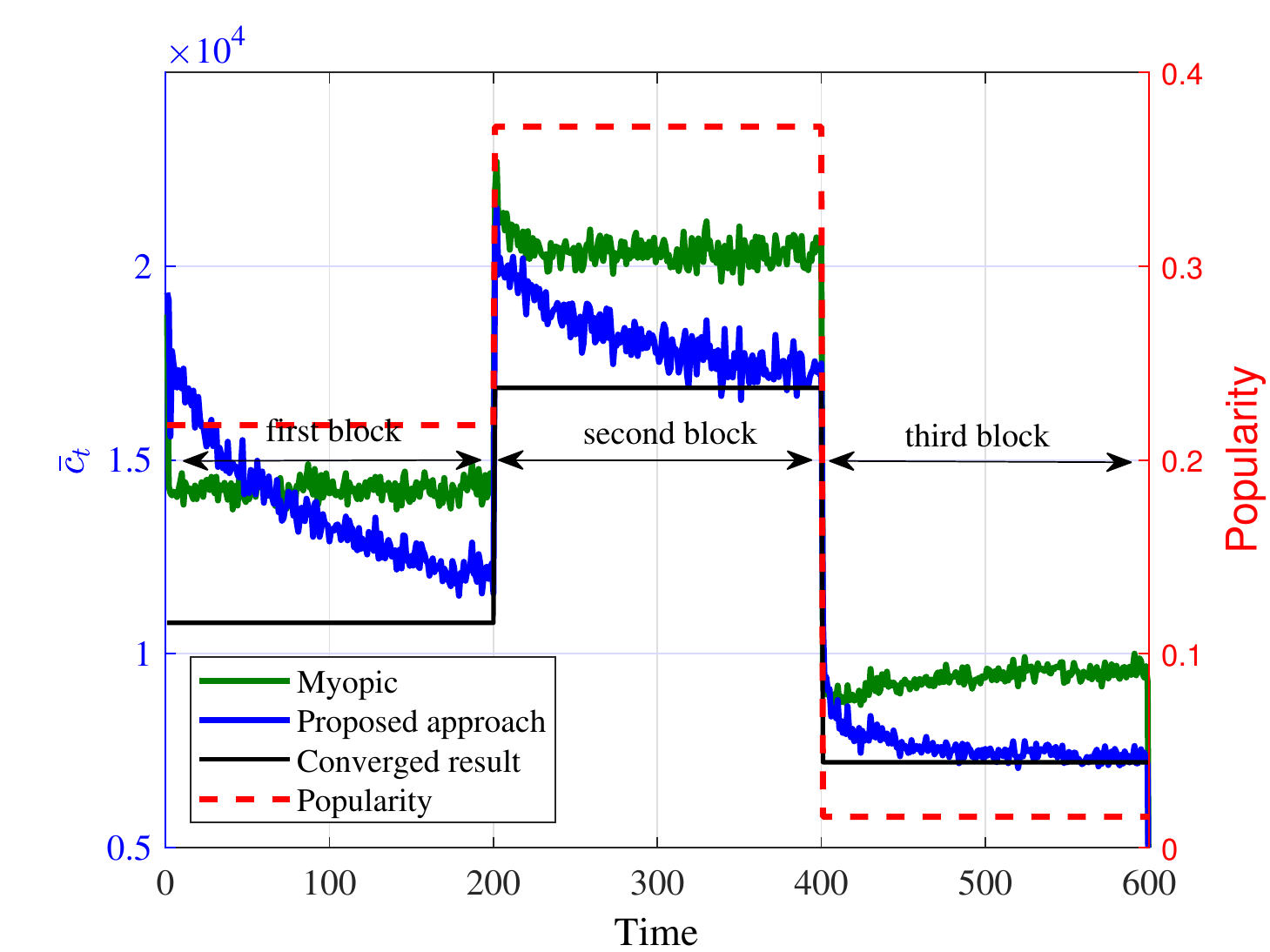}
	\caption{Averaged immediate cost over $1000$ realizations in a non-stationary setting, and a sample from popularities. }
	\label{result7}
\end{figure}

In this section, we numerically assess the  performance of the proposed approaches for learning optimal fetch-cache decisions. Two sets of numerical tests are provided. In the first set, summarized in Figs~\ref{fig:result1}-\ref{fig:result4}, the performance of the value iteration-based scheme in Alg. 1 is evaluated, and in the second set,  summarized in Figs.~\ref{result6}-\ref{result7},  the performance of the $Q$-learning solver is investigated. In both sets, the cache and fetch cost parameters are drawn with equal probability from a finite number of values, where the mean is $\bar{\rho}^f$ and  $\bar{\lambda}^f$, respectively. Furthermore, the request variable $r^f$ is modeled as a Bernoulli random variable with mean $p^f$, whose value indicates the popularity of file $f$. 

In the first set, it is assumed that $ p^f$ as well as the distribution of $\rho^f, \lambda^f$, are known a priori. Simulations are carried out for a content of unit size, and can be readily extended to files of different sizes. To help readability, we drop the superscript $f$ in this section.

Fig.~\ref{fig:result1} plots the sum average cost $\bar{\mathcal{C}}$ versus $\bar{\rho}$ for different values of $\bar{\lambda}$ and $p$. The fetching cost is set to $\bar{\lambda} \in \left\{43, 45, 50, 58\right\}$ for two different values of popularity $p  \in \{ 0.3,0.5\}$.  As depicted, higher values of $\bar{\rho},\bar{\lambda},p$ generally lead to a higher average cost. In particular, when $\bar{\rho}\ll \bar{\lambda}$, caching is considerably cheaper than fetching, thus setting $a_t=1$ is optimal for most $t$. As a consequence, the total cost linearly increases with $\bar{\rho}$ as most requests are met via cached contents rather than fetching. Interestingly, if $\bar{\rho}$ keeps increasing, the aggregate cost gradually saturates and does not grow anymore. The reason behind this  observation is the fact that, for very high values of $\bar{\rho}$, fetching becomes the optimal decision for meeting most file requests and, hence, the aggregate cost no longer depends on $\bar{\rho}$. While this behavior occurs for the two values of $p$, we observe that for the smallest one, the saturation is more abrupt and takes place at a lower  $\bar{\rho}$. The intuition in this case is that for lower popularity values, the file is requested less frequently, thus the caching cost aggregated over a (long) period of time often exceeds the ``reward'' obtained when (infrequent) requests are served by the local cache. As a consequence, fetching in the infrequent case of $r_t=1$ incurs less cost than the caching cost aggregated over time.

To corroborate these findings, Fig.~\ref{fig:result2} depicts the sum average cost versus $p$ for different values of  $\bar \rho$ and $\bar \lambda$. The results show that for large values of $\bar{\rho}$, fetching is the optimal action, resulting in a linear increase in the total cost as $p$ increases. In contrast, for small values of $\bar{\rho}$, caching is chosen more frequently, resulting in a sub-linear cost growth.
\hspace{0.1cm}
\hspace{0.1cm}
\hspace{0.1cm}
\vspace{0.1in}

To investigate the  caching-versus-fetching trade-off for a broader range of $\bar{\rho}$ and $\bar{\lambda}$, let us define the \textit{caching ratio} as the aggregated number of positive caching decisions (those for which $a_t=1$) divided by  the total number of decisions. Fig. \ref{fig:result3} plots this ratio for different values of $(\bar{\rho},\bar\lambda)$ and fixed $p=0.5$. As the plot demonstrates, when $\bar \rho$ is small and $\bar \lambda$ is large, files are cached almost all the time, with the caching ratio decreasing (non-symmetrically) as $\bar{\rho}$ increases and $\bar{\lambda}$ decreases.

Finally,   Fig.~\ref{fig:result4} compares the performance of the proposed DP-based strategy with that of a myopic one.  The myopic policy sets $a_t\!=\!1$ if $\lambda_t\!>\!\rho_t$ and the content is locally available (either because $w_t\!=\!1$ or because $s_t\!=\!1$), and sets $a_t\!=\!0$ otherwise.
The results indicate that the proposed strategy outperforms the myopic one for all values of $\bar{\rho},\bar{\lambda},p$ and $\gamma$.

In the second set of tests, the performance  of the online Q-learning solvers is investigated. As explained in Section \ref{Sec:DP_Formulation}, under the assumption that the underlying distributions are stationary, the performance of the Q-learning solver should converge to the optimal one found through the value iteration algorithm. Corroborating this statement, Fig.~\ref{result6} plots the sum average cost $\bar {\mathcal C}$ versus $\bar \rho$ of both the marginalized value iteration and the Q-learning solver, with $\bar \lambda \in \left\{29,36,44\right\}$ and $p \in \left\{0.3,0.5\right\}$. The solid lines are obtained when assuming a priori knowledge of the distributions and then running the marginalized value iteration algorithm; the results and analysis are similar to the ones reported for Fig.~\ref{fig:result1}. The dashed curves however, are found by assuming unknown distributions and running the 
Q-learning solver. Sum average cost is reported after first $1000$ iterations. As the plot suggests, despite the lack of a priori knowledge on the distributions, the Q-learning solver is able to  find the optimal decision making rule. As a result, it yields the same sum average cost as that of value-iteration under known distributions. 	

The last experiment investigates the impact of the instantaneous cache capacity constraint in C$4$ as well as non-stationary distributions for popularities and costs.  To this end, 1,000 different realizations (trajectories) of the random state processes are drawn, each of length $T=600$. For every realization, the cost $c_t$ [cf. \eqref{Sum_cost}] at each and every time instant is found, and the cost trajectory is averaged across the 1,000 realizations. Specifically, let $c_t^i$ denote the $i$th realization cost at time $t$, and define the averaged cost trajectory as ${\bar c}_t := \frac{1}{1000} \sum_{i=1}^{1000} c^i_t$. Fig.~\ref{result7} reports the average trajectory of ${\bar c}_t$ in a setup where the total number of files is set to $F = 500$, the file sizes are drawn uniformly at random from the interval $[1,100]$, and the total cache capacity is set to  $40\%$ of the aggregate file size. Adopted parameters for the MQ-learning solver are set to  $\beta = 0.3,$ and $ \epsilon = 0.01$. Three blocks of iterations are shown in the figure, where in each block a specific distribution of popularities and costs are considered. For instance, the dashed line shows the popularity of a specific file in one of the realizations, where in the fist block $p = 0.23$, in the second block $p = 0.37$, and in the third one  $p = 0.01$. The cost parameters have means $\bar \lambda = 44, \bar \rho = 2$,  $\bar \lambda = 40,\bar \rho = 5$, and $\bar \lambda = 38,\bar \rho = 2$ in the consecutive blocks, respectively. As this plot suggests, the MQ-learning algorithm incurs large costs during the first few iterations. Then, it gradually adapts to the file popularities and cost distributions, and learns how to make optimal fetch-cache decisions, decreasing progressively the cost in each of the blocks. To better understand the behavior of the algorithm and assess its performance, we compare it with that of a myopic policy and the stationary policy whose costs are represented using a green and black line, respectively. During the first iterations, when the MQ-learning algorithm has not adapted to the distribution of pertinent parameters, the myopic policy performs better. However, as the learning proceeds, the MQ-learning starts to make more precise decisions and, remarkably, in a couple of hundreds of iterations it is able to perform very close to the optimal policy.

\section{Conclusions}
\label{Sec_Conclusion}
A generic setup where a caching unit makes sequential fetch-cache decisions based on dynamic prices and user requests was investigated. Critical constraints were identified, the aggregated cost across files and time instants was formed, and the optimal adaptive caching was then formulated as a stochastic optimization problem.  Due to the effects of the current cache decisions on future costs, the problem was cast as a dynamic program. To address the inherent functional estimation problem that arises in this type of programs, while leveraging the underlying problem structure, several computationally efficient algorithms were developed, including off-line (batch) approaches, as well as online (stochastic) approaches based on Q-learning. The last part of the paper was devoted to dynamic pricing mechanisms that allowed handling constraints both in the storage capacity of the cache memory, as well as on the back-haul transmission link  connecting the caching unit with the cloud.

\bibliographystyle{IEEEbib}
\bibliography{biblio}

\end{document}